\documentclass[fleqn,usenatbib]{mnras}
\usepackage{newtxtext,newtxmath}
\usepackage[T1]{fontenc}
\usepackage{ae,aecompl}

\usepackage{color, soul, ulem,graphicx,commath}

\newcommand{\etor}{\langle B_{\rm tor}^2 \rangle}

\newcommand{\epol}{\langle B_{\rm pol}^2 \rangle}

\newcommand{\plm}{P_{\ell m}}
\newcommand{\lm}{_{\ell m}}
\newcommand{\tp}{(\theta, \varphi)}

\newcommand{\lmax}{\ell_{\rm max}}
\newcommand{\bv}{\langle |B_{V}| \rangle}
\newcommand{\bi}{\langle |B_{I}| \rangle}

\renewcommand{\l}{\ell}

\title[The magnetic field vector of the Sun-as-a-star. II]{The magnetic field vector of the Sun-as-a-star. II. Evolution of the large-scale vector field through  activity cycle 24}
\author[Vidotto, Lehmann, Jardine, Pevtsov]{A.~A.~Vidotto$^{1}$\thanks{E-mail: Aline.Vidotto@tcd.ie}, L.~T.~Lehmann$^{2}$, M. Jardine$^{2}$, A.~A.~Pevtsov$^{3}$
\\ 
$^{1}$School of Physics, Trinity College Dublin, the University of Dublin, Dublin-2, Ireland\\
$^{2}$  SUPA, School of Physics and Astronomy, University of St Andrews, North Haugh, St Andrews KY16 9SS, UK \\
$^{3}$ National Solar Observatory,  3665 Discovery Drive, 3rd Floor, Boulder, CO 80303 USA}

\date{Accepted XXX. Received YYY; in original form ZZZ}

\pubyear{2018}

\begin{document}
\label{firstpage}
\pagerange{\pageref{firstpage}--\pageref{lastpage}}
\maketitle

\begin{abstract}
In the present work, we investigate how the large-scale magnetic field of the Sun, in its three vector components, has evolved during most of cycle 24, from 2010 Jan to 2018 Apr. To filter out the small-scale field of the Sun, present in high-resolution synoptic maps, we use a spherical harmonic decomposition method, which decomposes the solar field in multipoles with different $\l$ degrees. By summing together the low-$\l$ multipoles, we reconstruct the large-scale field at a resolution similar to observed stellar magnetic fields, which allows the direct comparison between solar and stellar magnetic maps. During cycle 24, the `Sun-as-a-star' magnetic field  shows a polarity reversal in the radial and meridional components, but not in the azimuthal component. The large-scale solar field  remains mainly poloidal with $\gtrsim 70\%$ of its energy contained in the poloidal component. During its evolution, the large-scale field is more axisymmetric and more poloidal when near minima in sunspot numbers, and with a larger intensity near maximum. There is a correlation between  toroidal energy and sunspot number, which indicates that spot fields are major contributors to the toroidal large-scale energy of the Sun. The solar large-scale magnetic properties fit smoothly with observational trends of stellar magnetism reported in See et al. The toroidal ($\etor$) and poloidal ($\epol$) energies are related as $\etor \propto \epol ^{1.38\pm 0.04}$. Similar to the stellar sample, the large-scale field of the Sun shows a lack of  toroidal non-axisymmetric field. 
 \end{abstract}

\section{Introduction}
One of the most successful techniques for imaging stellar magnetism is the Zeeman Doppler Imaging (ZDI) technique. By analysing a time-series of circularly polarised spectra (Stokes V), one is able to recover information about the large-scale magnetic topology, including its intensity, orientation and how the magnetic field is distributed over the stellar surface \citep{1997A&A...326.1135D}. As a result, the ZDI technique reconstructs the vector magnetic field through the stellar surface, providing surface distributions of all three vector components of stellar magnetism (radial, meridional and azimuthal).

Recently,  synoptic maps of three vector components of the solar magnetic field have also started to be produced  \citep{2013ApJ...772...52G}. These maps are derived from full-disk vector magnetograms, e.g., observed by the Synoptic Optical Long-term Investigations of the Sun (SOLIS, \citealt{2003SPIE.4853..194K}) at the National Solar Observatory and with the Helioseismic and Magnetic Imager (HMI, \citealt{2012SoPh..275....3P,2014SoPh..289.3483H}) at the Solar Dynamics Observatory. The new vector synoptic fields allow us to investigate the meridional (North-South) and azimuthal (East-West) components of the solar magnetic field, in addition to the well-studied radial (up-down relative to local surface of the Sun) component. Furthermore, we can now compare vector maps of the Sun to ZDI maps, to investigate whether the solar magnetic field is typical among other low-mass stars \citep{2016MNRAS.459.1533V, 2017MNRAS.466L..24L, 2018MNRAS.tmp.1193L}. 

However, the difference in resolution between solar and stellar maps poses a problem in such a direct comparison. 
Stellar magnetic maps have much lower resolution than solar maps. In particular, the small-scale structure observed at the surface of the Sun (e.g., sunspots, active regions) cannot be seen by ZDI, due to flux cancelation within an element of resolution \citep{2010MNRAS.404..101J, 2011MNRAS.410.2472A, 2014MNRAS.439.2122L}. ZDI can only see the  large-scale magnetic field (i.e., large-scale components such as the dipole, quadrupole and further combinations of low-degree multipoles). However, in the Sun, the magnetic flux of solar active regions dominates, being orders of magnitude larger than the magnetic flux of the large-scale magnetic field.  To provide a direct comparison between solar synoptic maps and stellar ZDI maps, one needs to filter out the small-scale field (i.e., the finely resolved magnetic features) of the solar observations. This method consists of decomposing the high-resolution solar magnetic maps in spherical harmonics and it was used in the context of the line-of-sight solar observations \citep{2012ApJ...757...96D,2013ApJ...768..162P} and recently in the context of vector fields \citep{2016MNRAS.459.1533V, 2017MNRAS.466L..24L}. In other words, the magnetic field is considered to be a superposition of magnetic multipoles with different degrees $\l$. Each multipole has a different intensity and angle with respect to the polar axis. The smallest multipole $\l$ degrees represent the largest-scale components of the magnetic field, e.g., $\l=1$ for the dipole, $\l=2$ for the quadrupole, $\l=3$ for the octupole and so on, while higher values of $\l$ represent increasingly smaller scales at the surface of the Sun.

Once the spherical harmonics decomposition is done, one can then  add individual multipoles up to a maximum degree $\lmax$ of interest. For the case where one wants to reconstruct the large-scale field of the Sun and compare it to ZDI maps, $\lmax$ should be chosen to be consistent with values used in ZDI studies. Typically, a ZDI map is able to reconstruct multipoles up to degrees $5$ to $15$, depending on the resolution of the observations. The method we use here is consistent with the magnetic field description adopted in several ZDI studies \citep[e.g.][]{2006MNRAS.370..629D, 2016MNRAS.457..580F, 2016A&A...593A..35R}, so it is ideal for comparing large-scale structure of the solar field with stellar magnetism. 

An alternative way to reconstruct the large-scale field of the Sun and other stars is to use magnetic field data (observational or numerical), add  noise representative of stellar observations, and conduct a ZDI reconstruction of the Sun-as-a-star magnetic field. Using this method, Lehmann et al (in prep. a) conducted a series of ZDI reconstructions on the simulated data of solar magnetic fields, based on the 3D non-potential flux transport simulations from \citet{2016MNRAS.456.3624G}.  They demonstrated that ZDI recovers the simulated large-scale solar magnetic field provided the Sun is observed with a higher ($i\gtrsim 60^{\rm o}$) inclination angle, as is the case of most stars with ZDI maps. If the star is observed at lower inclinations (more pole-on), $v\sin(i)$ becomes small, worsening the resolution in ZDI. In this case, ZDI cannot distinguish between hemispheres. They also showed that our method of harmonics filtering recovers a similar field structure to what would be recovered if one would do a full ZDI field reconstruction of the solar magnetic field (i.e., from a time-series of Stokes V profiles). The harmonics filtering method we use in the present study is significantly less time-consuming than the ZDI reconstruction.

The goal of the present study is to investigate the evolution of the large-scale  field of the Sun in its three vector components and compare it with stellar magnetic properties derived in ZDI studies. For that, we derive the spherical harmonic coefficients from solar synoptic maps of the vector field, filter out the high $\l$-degrees (i.e., removing the small-scale field and keeping the degrees that represent the large-scale field) and reconstruct only the large-scale field component. We do this for a series of HMI solar magnetic maps from 2010~Jan to 2018~April\footnote{{The HMI maps, downloaded as fits files from \url{http://jsoc.stanford.edu/ajax/lookdata.html?ds=hmi.b_synoptic},  have a resolution of $3600 \times 1440$ pixel$^2$, which corresponds to  360 degrees in solar longitude and $\pm 1$ in sine of solar latitude.}} \citep{liu}. This period roughly coincides with the beginning of solar cycle 24 and is approaching its end -- at the two extrema, the number of spots observed at the surface of the Sun is quite low (14 and 3, respectively). In contrast, the maximum sunspot number observed during this interval is $\sim 114$, around mid-2014 (Figure \ref{fig.SN}). The large-scale field of the Sun evolves during its activity cycle. While many studies investigated this evolution using the radial component of the solar magnetic field  \citep[e.g.][]{2003JGRA..108.1035S,2012ApJ...757...96D, 2013ApJ...768..162P}, there is no study investigating the evolution of the large-scale meridional and azimuthal components nor the toroidal and poloidal fields. By studying the large-scale magnetic evolution of the solar cycle, we  gain a better understanding of activity and magnetic cycles in other stars. With long-term monitoring, the evolution of the magnetic field has been observed in several stars now \citep[e.g.][]{2012A&A...540A.138M, 2015A&A...573A..17B, 2016A&A...593A..35R, 2017MNRAS.471.1246F}, and magnetic cycles have been confirmed in a few targets such as $\tau$ Boo \citep{2009MNRAS.398.1383F, 2013MNRAS.435.1451F, 2016MNRAS.459.4325M, jeffers}, 61 Cyg A \citep{2016A&A...594A..29B}, and $\epsilon$ Eridani \citep{2014A&A...569A..79J, 2017MNRAS.471L..96J}.  
 
This paper is divided as follows: Section \ref{sec.mathematical} overviews the method we use to decompose the magnetic field vector using spherical harmonics and the subsequent reconstruction of the large-scale magnetic field. Sections \ref{sec.reversals}, \ref{sec.poltor} and \ref{sec.evolution} present our results: the polarity reversals in the large-scale vector solar field during cycle 24, how the magnetic energy is distributed in the poloidal and toroidal components and the magnetic evolution through cycle 24. Section \ref{sec.zdi} puts the results of the large-scale solar field in the context of stellar magnetic maps.  Section \ref{sec.conclusions} presents a summary of our results and our conclusions.

\section{Magnetic field decomposition using vector spherical harmonics}\label{sec.mathematical}

\subsection{Mathematical formalism}
The radial $B_r$, meridional $B_\theta$ and azimuthal $B_\varphi$ components of the magnetic field are expressed in terms of spherical harmonics as
\begin{equation}\label{eq.br}
B_r \tp =    \sum\lm \alpha\lm P\lm e^{im\varphi} \, ,
\end{equation}
\begin{equation}\label{eq.btheta}
B_\theta \tp =   \sum\lm \beta\lm \frac{1}{l+1} \frac{\dif \plm}{\dif \theta} e^{im\varphi}  + \gamma\lm  \frac{ im \plm  e^{im\varphi}}{(l+1) \sin\theta}  \, ,
\end{equation}
\begin{equation}\label{eq.bphi}
B_\varphi \tp =- \sum\lm \beta\lm  \frac{ im \plm  e^{im\varphi}}{(l+1) \sin\theta} - \gamma\lm \frac{1}{l+1} \frac{\dif \plm}{\dif \theta} e^{im\varphi}   \, , 
\end{equation}
where $ \plm \equiv \plm (\cos \theta)$ is  the associated Legendre polynomial of degree $\l$ and order $m$. The sums are performed over $1\leq \ell \leq \ell_{\rm max}$  and  $|m| \leq \ell$, where $\ell_{\rm max}$ is the maximum degree of the spherical harmonic decomposition.

To filter out the small-scale magnetic fields, we proceed as follows: from the observed surface distributions of $B_r$,  $B_\theta $ and $B_\varphi$, we calculate the  spherical harmonic coefficients  $\alpha\lm$, $\beta\lm$ and $\gamma\lm$ by inverting Equations (\ref{eq.br}) to (\ref{eq.bphi}), as presented in \citet{2016MNRAS.459.1533V}.  Once the coefficients are computed, we use them in Equations (\ref{eq.br}) to (\ref{eq.bphi}), limiting the sums to a maximum value of $\ell_{\rm max}$.  Truncating the sums to a higher value for $\ell_{\rm max}$ implies a higher resolution in the field reconstruction. If $\ell_{\rm max}$ is sufficiently large, the high-resolution observed maps are recovered. In the example presented in \citet{2016MNRAS.459.1533V}, a value of $\ell_{\rm max}=150$ was demonstrate to recover the original observed magnetic field maps.  
Because the goal of this paper is to investigate the large-scale magnetic field vector of the Sun (i.e., the Sun as a star), we limit the sums to a maximum value of $\ell_{\rm max}=5$. This way, we keep only the magnetic field components up to $\ell_{\rm max}=5$, filtering out  magnetic multipoles with degree  $\l>5$. 
This is motivated by the results of Lehmann et al (in prep a), who showed that ZDI recovers only significant magnetic energies for $\l$-degrees up to $\l = 5$ for simulated solar-like rotators. Fast rotating stars typically allow a higher value of $\lmax \sim 15$ to be adopted in the ZDI field reconstruction (some of our analysis in this paper adopts $\lmax=15$). This is due to a higher resolution in spectropolarimetric observations of fast rotators than that of slow, solar-like rotators.

In ZDI studies, the vector magnetic field is often decomposed in poloidal and toroidal components, with observations  showing that  stars can have significant surface toroidal fields \citep{2008MNRAS.388...80P, 2009ARA&A..47..333D, 2016MNRAS.462.4442S}. While most dynamo models enforce a zero toroidal field at the surface, \citet{2016ApJ...833L..22B} showed that it is possible for a dynamo model to produce a significant toroidal field. The terms with $\alpha\lm$ and $\beta\lm$ represent the poloidal part and the terms with $\gamma\lm$ the toroidal part of the field in Equations (\ref{eq.br}) to (\ref{eq.bphi}). Both components add such that $\vec{B}_{\rm pol} + \vec{B}_{\rm tor} = \vec{B}$. In some works, toroidal fields are used as synonyms to azimuthal field, but this is not the case in the present study. In the limit of a purely axisymmetric field ($m=0$), the toroidal field has only azimuthal component and the poloidal field only has radial and meridional components (i.e., it lies in meridian planes). In a general case, like ours, the toroidal field has meridional $\theta$ and azimuthal $\varphi$ components, and the poloidal field has all three components.

\subsection{Definitions used in this study}
{We denote $\vec{B}_\ell$ as the magnetic field strength at a specific $\ell$ degree.  The sums of individual $\vec{B}_\ell$  up to a degree $\lmax$ results in the total field strength $\vec{B}_\Sigma $. We define the magnetic field energy as the squared magnetic field averaged over the surface $ \langle B_x^2 \rangle = \frac{1}{4\pi}\int  \vec{B}_x\cdot \vec{B}_x \dif \Omega $, where $\Omega$ is the solid angle. The cumulative energy, i.e., when the magnetic field is summed over all $\ell$'s, is represented by $\langle B^2_\Sigma \rangle$, while the energy at a specific $\ell$ degree is represented by $ \langle B^2_\ell \rangle$. }

Analogously, the energy in the poloidal component is defined as $\langle B_{\rm pol}^2 \rangle = \frac{1}{4\pi}\int  \vec{B}_{\rm pol}\cdot \vec{B}_{\rm pol} \dif \Omega $ and the toroidal energy is defined in a similar way. To be more explicit, when discussing the cumulative energies in the poloidal field, we will use $\langle B^2_{{\rm pol},\Sigma} \rangle$ and for the energies in individual $\l$'s, we will use $\langle B^2_{{\rm pol},\ell} \rangle$. 
The fraction of poloidal fields is then $f_{{\rm pol}}= {\langle B_{{\rm pol},\Sigma}^2 \rangle}/{\langle B_{\Sigma}^2 \rangle}$ and $f_{{\rm pol}, \ell}= {\langle B_{{\rm pol},\ell}^2 \rangle}/{\langle B_{\ell}^2 \rangle}$, for cumulative and individual $\ell$ distributions, respectively.   The toroidal energies are defined in a similar fashion as above. The toroidal  fractions are $f_{\rm tor} = 1 - f_{\rm pol}$ or $f_{{\rm tor},\ell} = 1 - f_{{\rm pol}, \ell}$.
 
 Our analysis also takes into account the axisymmetry of the magnetic field. To calculate the axisymmetric component of the magnetic field vector, we only allow $m=0$ in the sums in Equations (\ref{eq.br}) to (\ref{eq.bphi}). The axisymmetric magnetic field energy is $\langle B_{\rm axi}^2 \rangle = \frac{1}{4\pi}\int  \vec{B}_{\rm axi}\cdot \vec{B}_{\rm axi} \dif \Omega $ and $f_{\rm axi}= {\langle B_{\rm axi}^2 \rangle}/{\langle B_\Sigma^2 \rangle}$. This definition is similar to that used in \citet{2015MNRAS.453.4301S} and \citet{2017MNRAS.466L..24L, 2018MNRAS.tmp.1193L}. In some other ZDI works, alternative definitions of the axisymmetric condition have been used \citep[e.g., $m < \ell/2$, ][]{2009MNRAS.398.1383F}.

\section{Polarity reversals in the large-scale field}\label{sec.reversals}
\begin{figure}
	\includegraphics[width=0.47\textwidth]{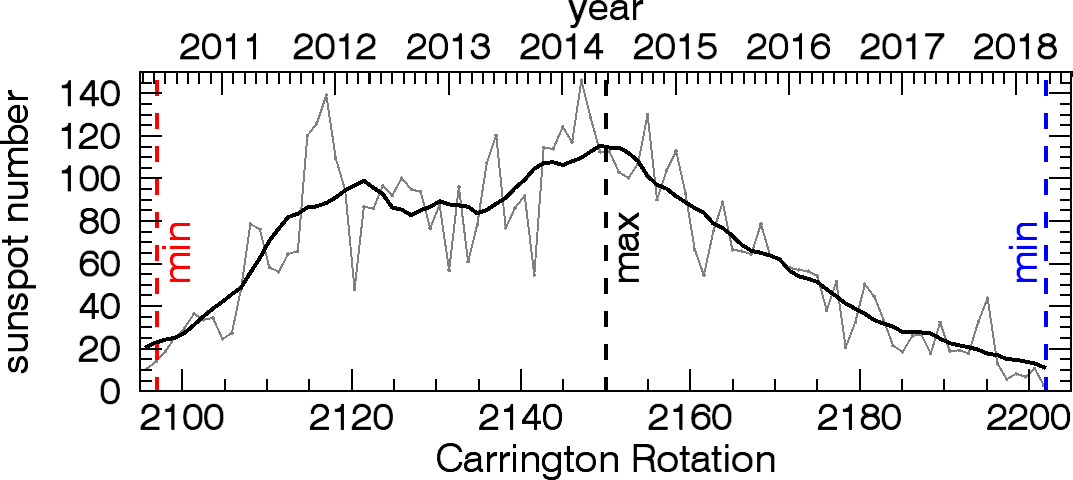}
\caption{Sunspot number (grey) and its smoothed values (black) as a function of time and Carrington Rotation (CR). The corresponding CR of the maps presented in Figure \ref{fig.sunmap} are shown by the dashed lines: near sunspot minima (red and blue with 14 and 3 spots, respectively) and maximum (black with $\sim 114$ spots).}\label{fig.SN}
\end{figure}

After calculating $\alpha\lm$, $\beta\lm$ and $\gamma\lm$, we then compute the large-scale field by restricting the sums in Equations (\ref{eq.br}) to (\ref{eq.bphi}) for $\ell\leq \ell_{\rm max} =5$. This process is done for all available maps, which, in the case of the vector synoptic maps from HMI, range from carrington rotations (CRs) CR2097 to CR2202. Due to their lack of data during one full CR, we removed maps CR2181 and CR2136 from our analysis, resulting in 104 analysed maps. The sunspot numbers (SSN) observed during this period is shown in Figure \ref{fig.SN}. The maps used in this study are part of cycle 24 and they range from nearly two consecutive minima in SSN.
  
\begin{figure*}
	\includegraphics[height=0.55\textwidth]{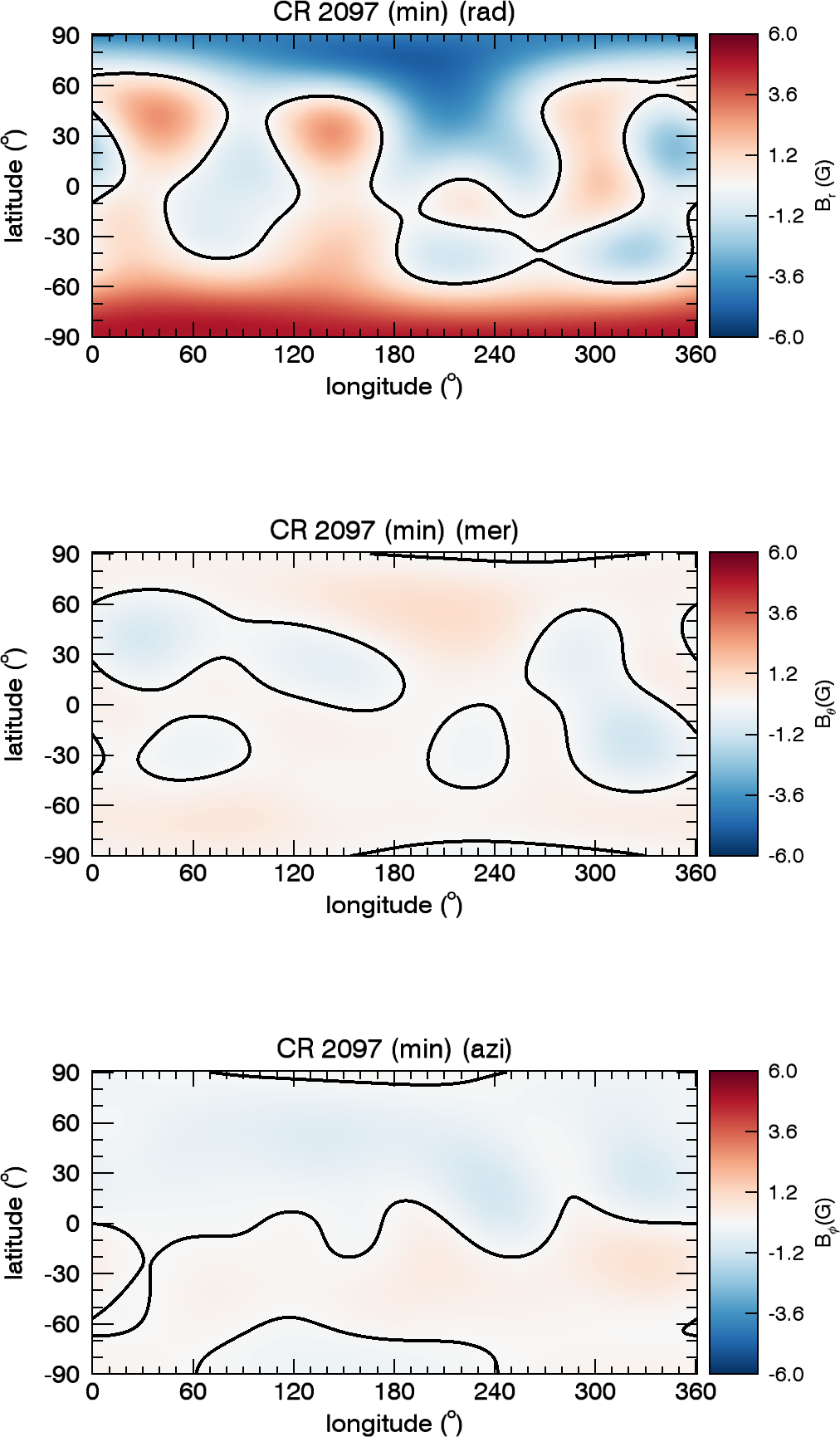}
	\includegraphics[height=0.55\textwidth]{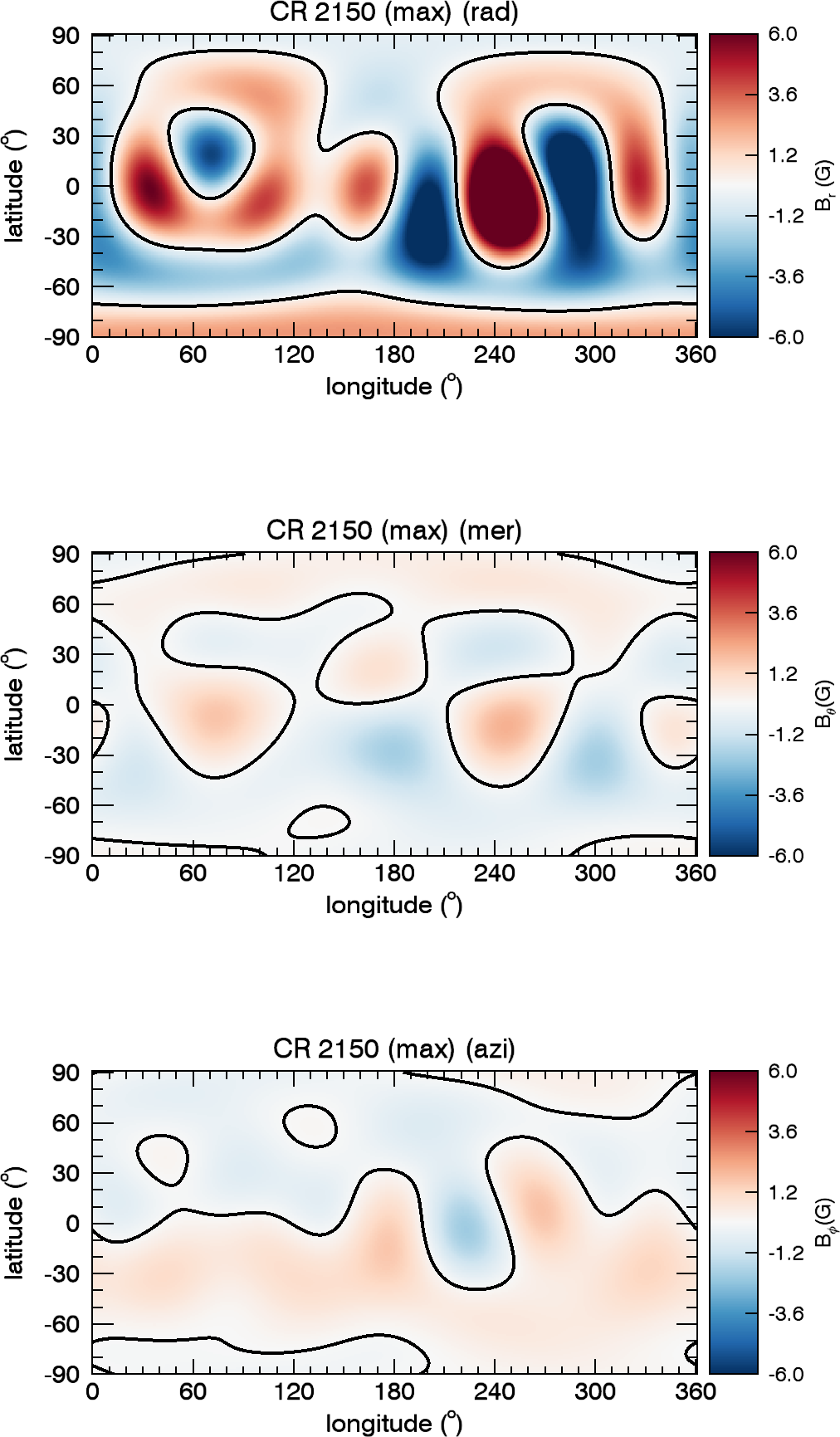}
	\includegraphics[height=0.55\textwidth]{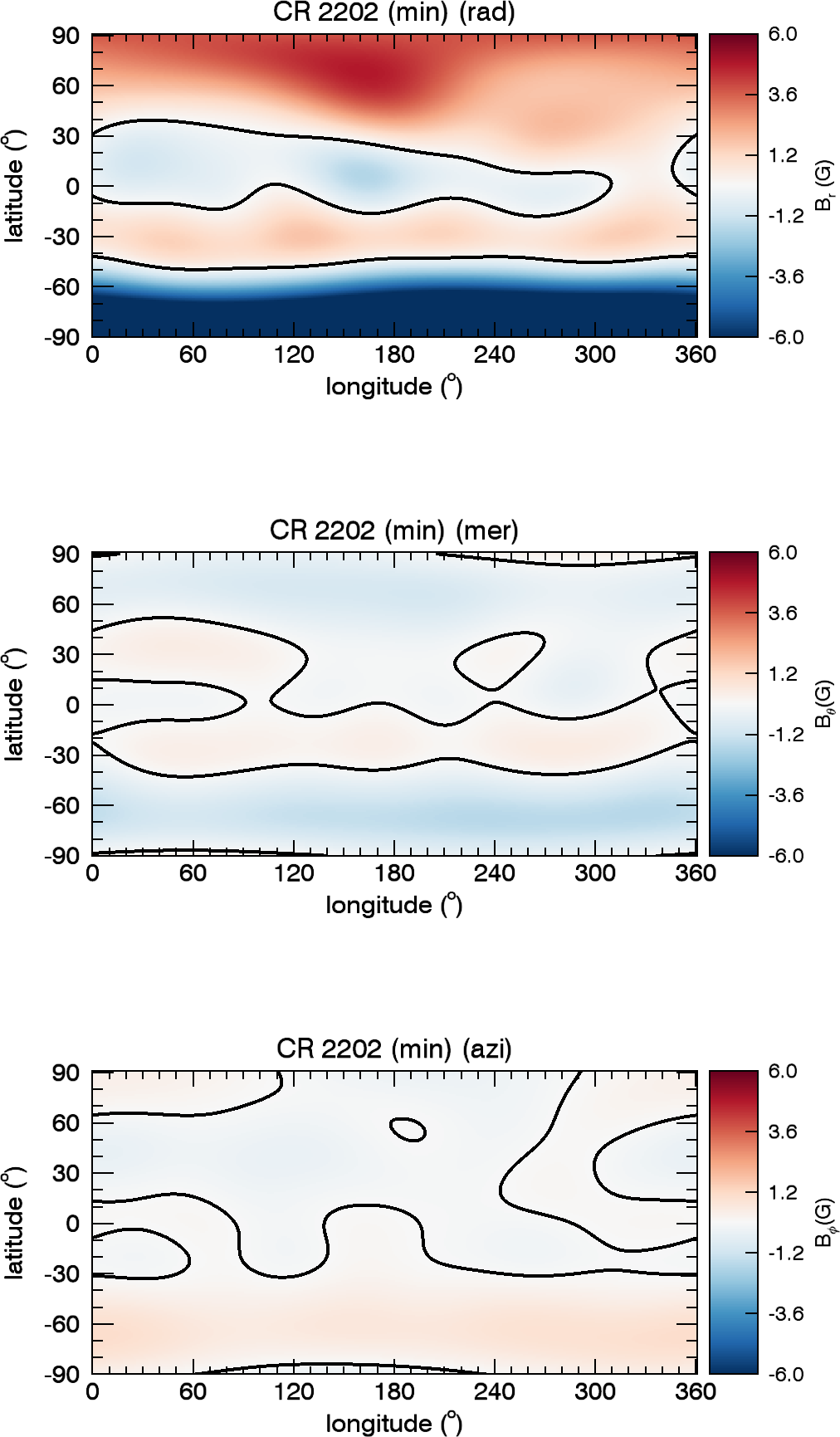}
\caption{Large-scale magnetic field reconstructed from high-resolution solar vector synoptic maps, obtained by restricting the spherical harmonics reconstruction up to a degree $\lmax = 5$, which is typical in stellar studies. From top to bottom: radial, meridional and azimuthal components. First and third column of maps show the field configuration of the Sun when it is nearly in sunspot minima (CR2097 and CR2202, respectively) and the middle column corresponds to when the Sun is near sunspot maximum (CR2150). The corresponding spherical harmonic coefficients are shown in the tables in Appendix~\ref{sec.coeff}.}\label{fig.sunmap}
\end{figure*}

Figure \ref{fig.sunmap} shows the large-scale magnetic field vector of the Sun at three different CRs: CR2097, CR2150 and CR2202. These CRs are represented  in Figure \ref{fig.SN} as dashed lines and they were chosen as to represent minimum, maximum and minimum in SSN, respectively. Figure \ref{fig.sunmap} shows that the polarities of the radial and meridional fields are different in the two consecutive minima, indicating a field reversal between CR2097 and CR2202. There is no reversal in the azimuthal component. A closer analysis shows that the azimuthal component is mainly from the toroidal part of the field, which does not reverse polarity throughout the 11-yr activity (sunspot) cycle. Because of this, the azimuthal field remains with roughly two polarity bands through the activity cycle: a positive azimuthal band in the southern hemisphere and a negative azimuthal band in the northern hemisphere. 

\begin{figure}
\centering
	\includegraphics[height=0.44\textwidth]{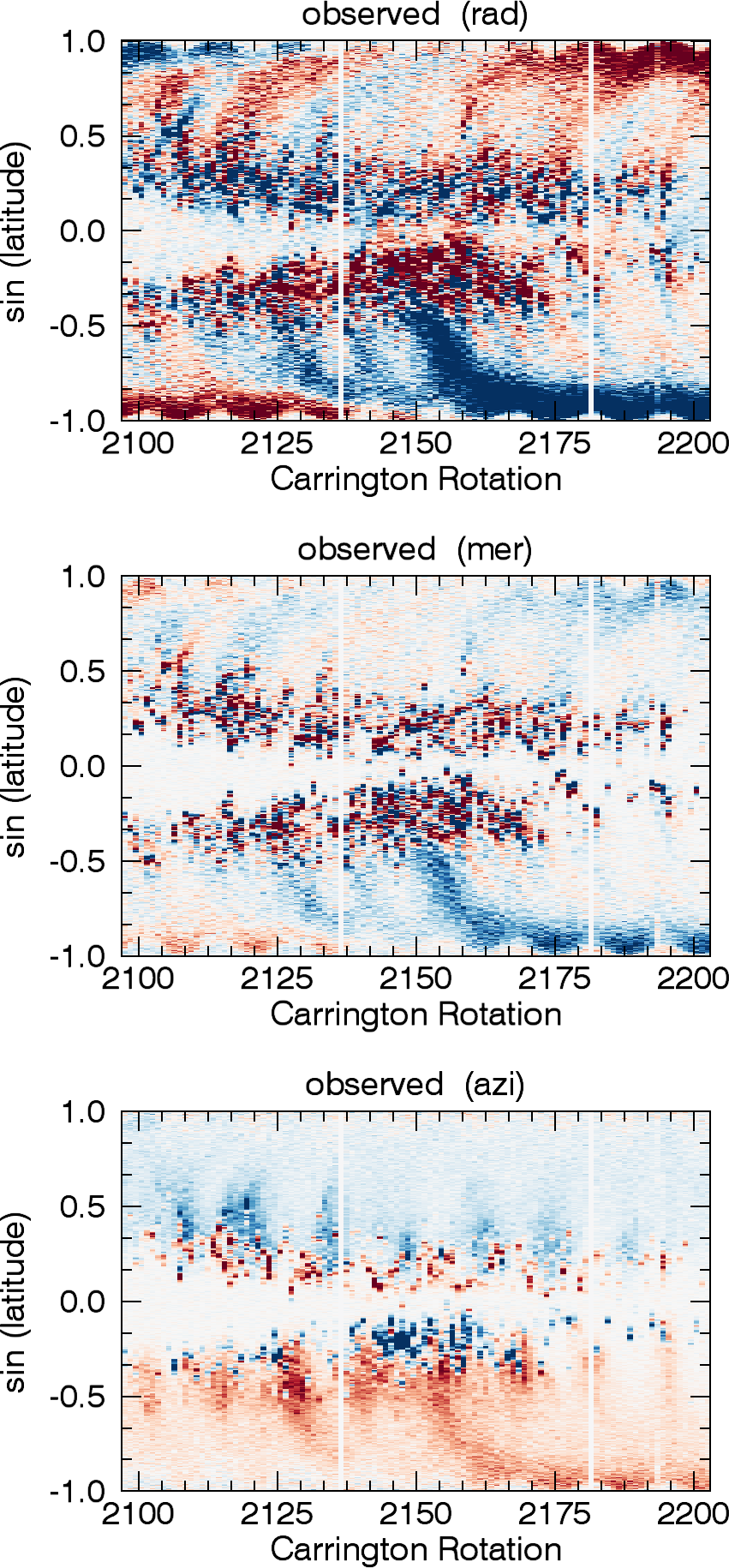}
	\includegraphics[height=0.44\textwidth]{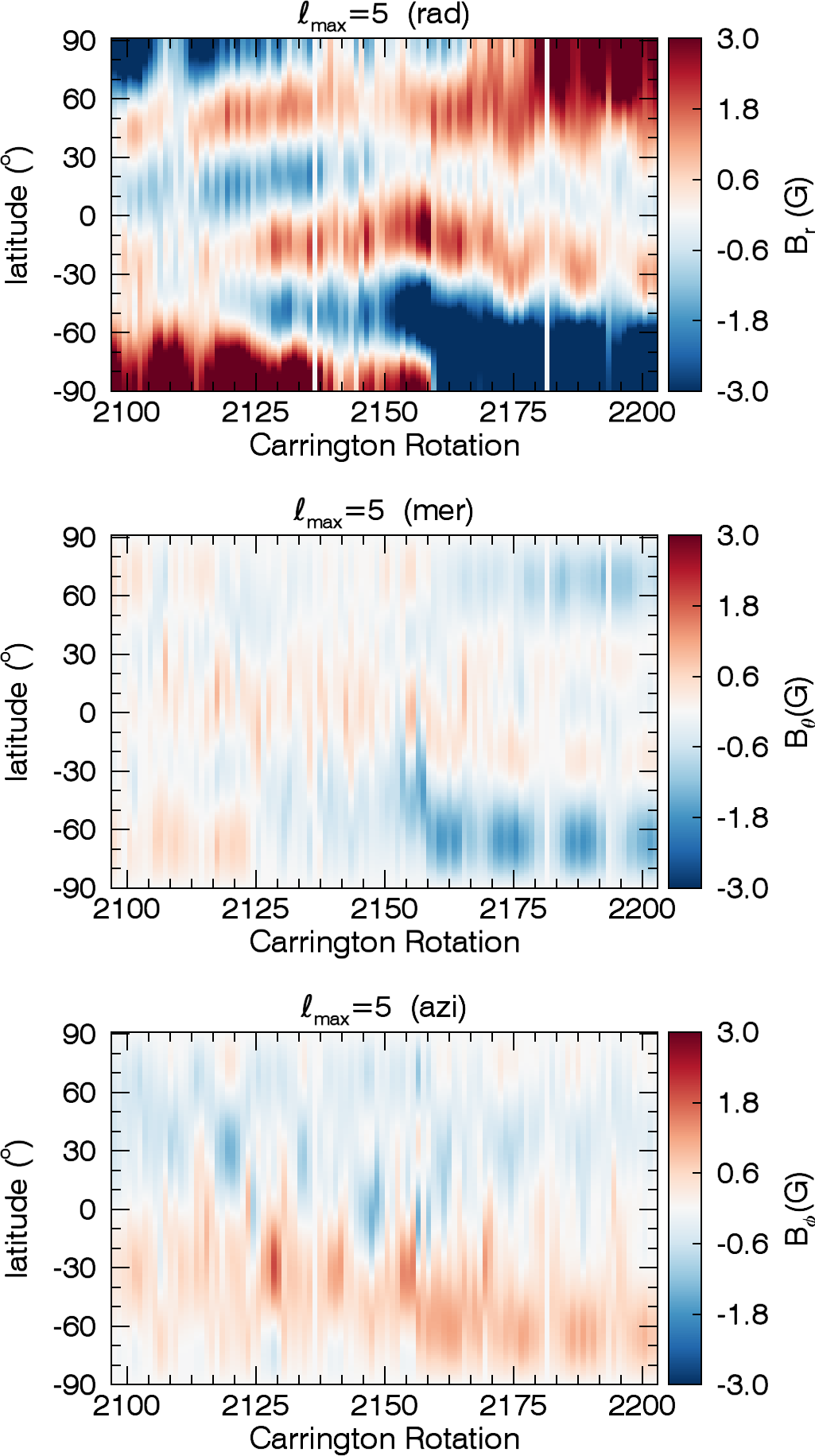}
\caption{Time-latitude diagrams of the observed (left) and large-scale (right) magnetic field of the Sun for $\lmax=5$. The $y$-axis is shown as sine latitude and latitude in the left and right panels, respectively. The wiggles seen in high latitudes are likely due to the inclination of the solar rotation axis with respect to the ecliptic. Note the field reverses polarity in $B_r$ and $B_\theta$ components (top two rows), but not yet in $B_\varphi$ (bottom row), which is expected to reverse sign at the beginning of next sunspot cycle.}\label{fig.butterfly}
\end{figure}

To further investigate field reversals, we average each component of the magnetic field over longitudinal bands and stack together the derived one-dimensional arrays to obtain a temporal evolution. The left panels in  Figure \ref{fig.butterfly} show the time-latitude diagram for the observed synoptic maps and the right panels for the large-scale magnetic field of the Sun for  $\lmax=5$ (note that the $y$-axis is  sine of latitude on the left panels). Missing maps were assigned a zero value in all their magnetic field components. We clearly see that the radial and meridional components reverse polarities in the total observed field (left) and large-scale components (right), but the azimuthal component remains mainly negative (positive) in the northern (southern) hemisphere. It is interesting to note also that the polarity of the active regions in the azimuthal component is opposed to the polarity of the magnetic field at higher latitudes (bottom left) at the same hemisphere. 
 
We believe that we still do not see the reversal in the azimuthal component because of the reduced timespan of maps, which do not yet probe the beginning of the next sunspot cycle (i.e., we do not have observations spanning the entire 22-yr magnetic cycle). As we show in Section \ref{sec.evolution}, $B_\varphi$, which is the dominant component forming the toroidal field,  is tracing the spot (active regions) fields. In a spatial average, $B_r$ of opposite polarity cancels out, however, because the majority of all active regions follows the Hale polarity orientation, $B_\varphi$ does not cancel out completely in a spatial averaging. If there is no systematic $B_\varphi$ component (e.g., due to noise) in the quiet Sun (i.e., outside of active regions), all $B_\varphi$ signal originates from bipolar structures of decaying active regions. Thus, we need to wait  for the beginning of the next sunspot cycle to be able to see the reversal in $B_\varphi$,  as the polarity orientation of the bipoles reverses sign at the beginning of each activity cycle. Therefore, our analysis shows that $B_\varphi$ has not reversed its polarity during cycle 24 and we expect a lag between the reversals in $B_r$ (which occurs after maximum of current cycle 24) and $B_\varphi$ (which will occur at the beginning of the next cycle 25).

Another feature we notice in Figure \ref{fig.butterfly} are the wiggles of the observed (left panels) and reconstructed (right panels) field, more clearly seen in the regions of high latitude. This modulation is due to the viewing angle of the solar poles due to the inclination of the solar rotation axis with respect to the ecliptic \citep{2015MNRAS.453L..69P}. The left panels in Figure \ref{fig.butterfly} for example show that the wiggles near the poles are in anti-phase between northern and southern hemispheres. At low latitudes, the effect is much smaller, and thus, the modulation is not that prominent. 

\section{Energy distributions in the poloidal and toroidal fields}\label{sec.poltor}
\begin{figure*}
	\includegraphics[width=0.47\textwidth]{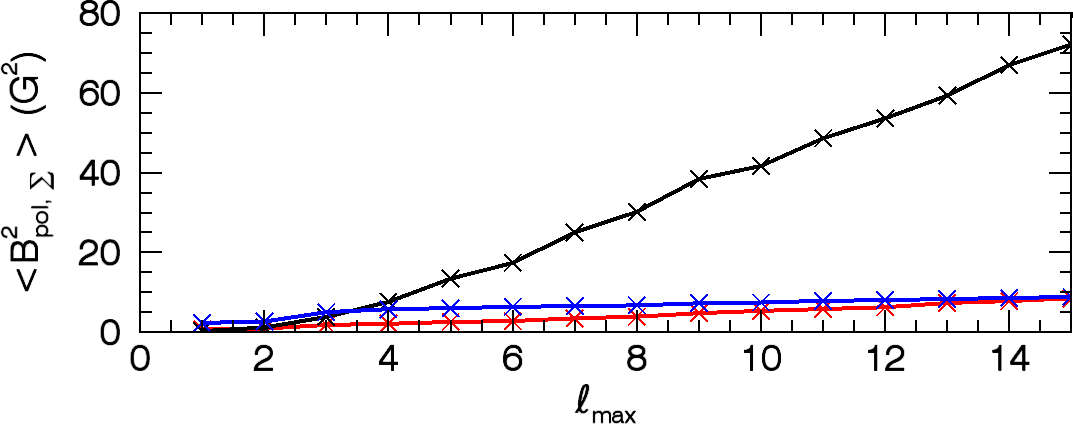}\,\,\,\,\,\,\,\,\,\,\,\,\,\,\,
	\includegraphics[width=0.47\textwidth]{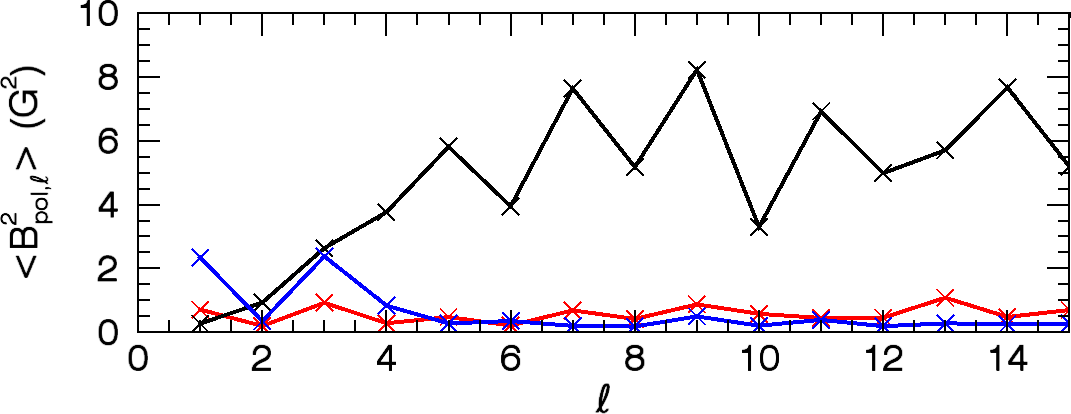} \\
 	\includegraphics[width=0.47\textwidth]{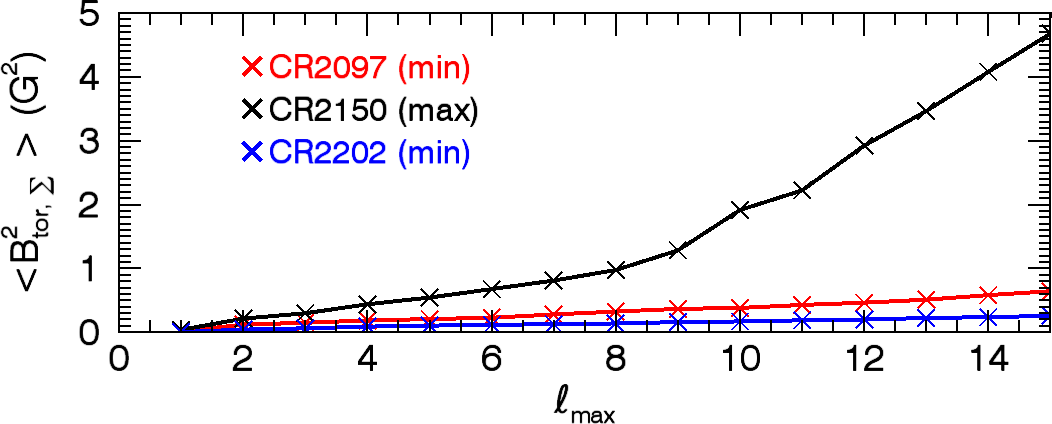}\,\,\,\,\,\,\,\,\,\,\,\,\,\,\,
	\includegraphics[width=0.47\textwidth]{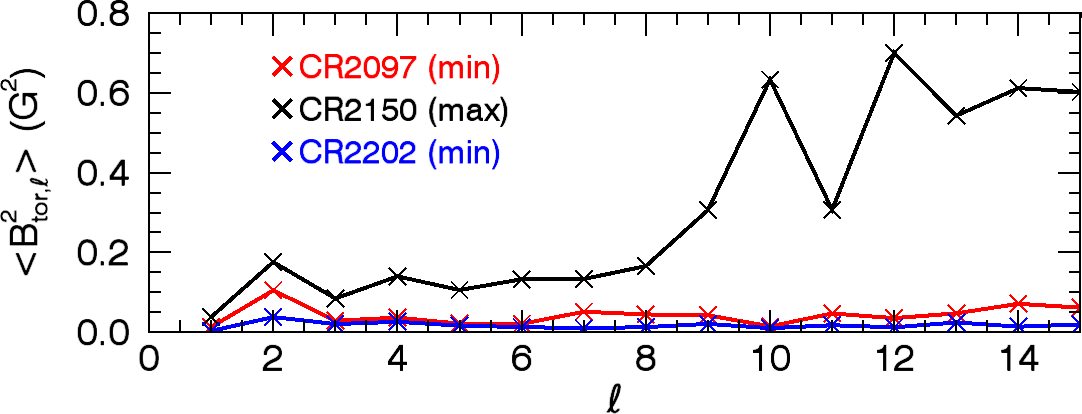}	 \\
 	\includegraphics[width=0.47\textwidth]{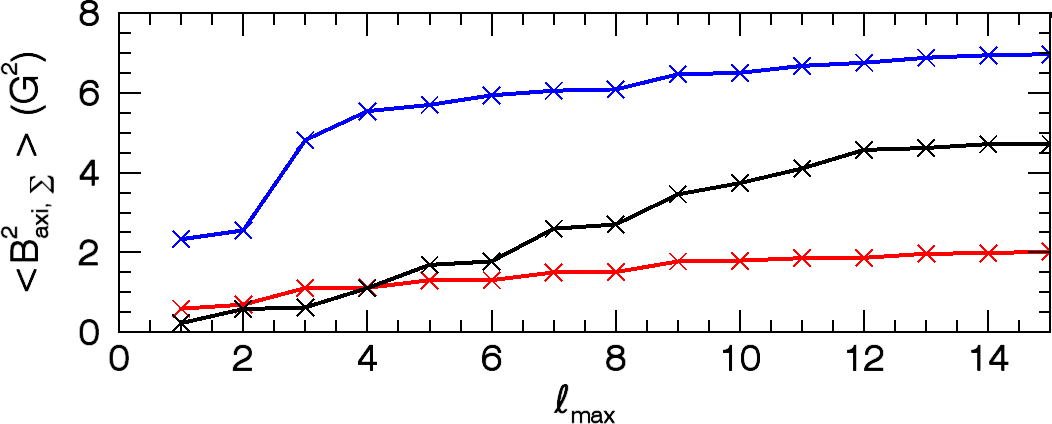}\,\,\,\,\,\,\,\,\,\,\,\,\,\,\,
	\includegraphics[width=0.47\textwidth]{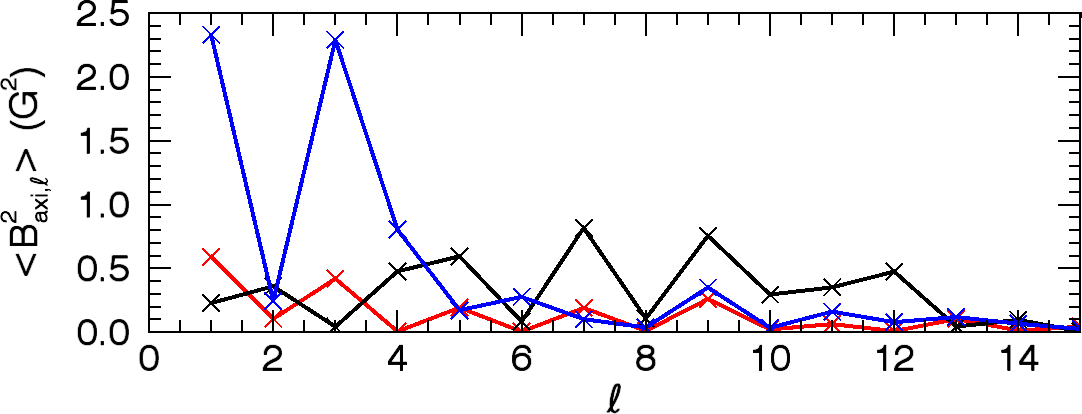}	
	\caption{Left panels:  Cumulative distribution of the magnetic energies as a function of $\ell_{\rm max}$  for different CR (min: red, max: black, min: blue). Right panels: Distribution of the magnetic energies contained in each $\ell$ degree for different CR.   {The top panels show poloidal energies, the middle panels toroidal energies and bottom panels show axisymmetric energies.}	}\label{fig.EpolEtor}
\end{figure*}

\begin{figure*}
	\includegraphics[width=0.9\textwidth]{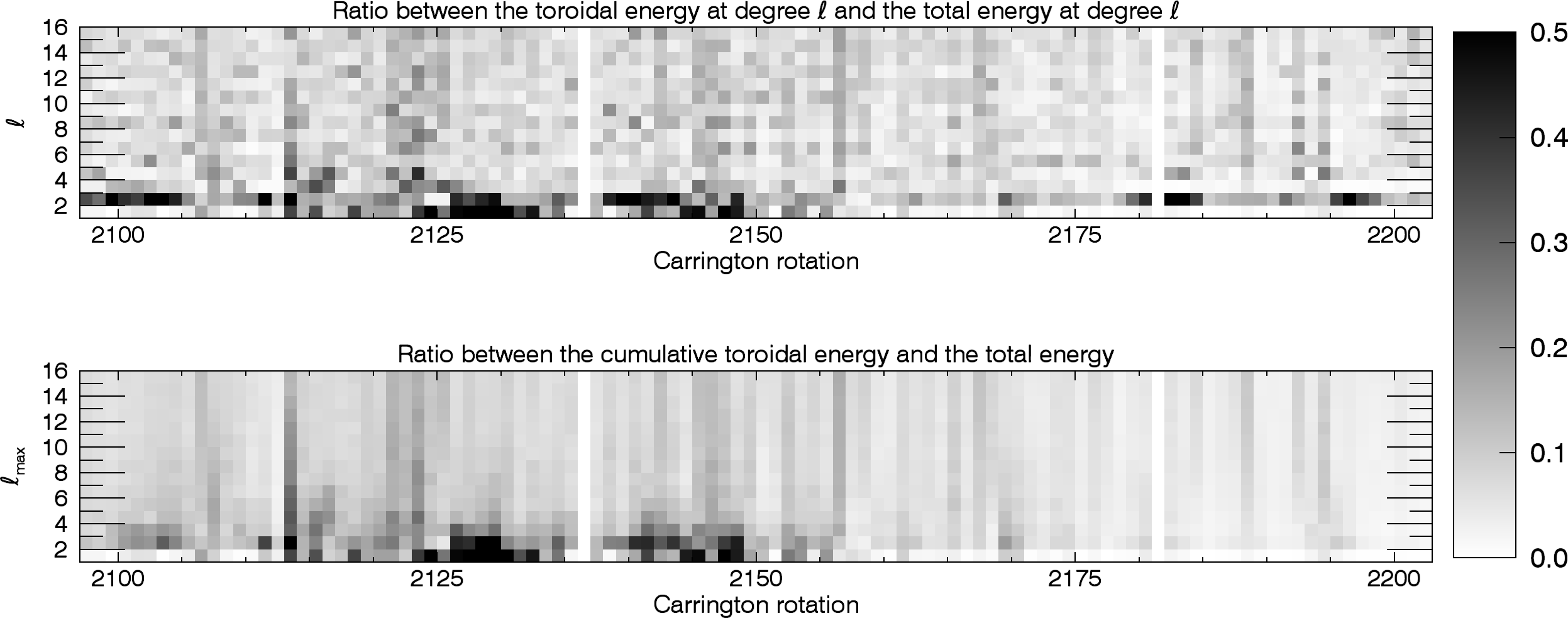}
\caption{{Top: Ratio between the  toroidal energy and the total energy   of the large-scale field of the Sun as a function of time. Energies are calculated at individual $\ell$ degree, i.e., $f_{{\rm tor}, \ell}=  {\langle B_{{\rm tor},\ell}^2 \rangle}/{\langle B_{\ell}^2 \rangle}$. Bottom: The same, but instead considering cumulative energies, i.e., $f_{\rm tor}={\langle B_{{\rm tor},\Sigma}^2 \rangle}/{\langle B_{\Sigma}^2 \rangle}$.} Note saturation of values above $0.5$.}\label{fig.fpol}
\end{figure*}

Figure \ref{fig.EpolEtor} shows the poloidal (top panels), toroidal (middle)  and {axisymmetric (bottom)} energies of the solar magnetic field at the same CRs plotted in Figure \ref{fig.sunmap}. The panels on the right show the energies at each individual degree $\l$, while the panels on the left show the cumulative sums (i.e., the sums of the energy per degree) up to a given $\ell_{\rm max}$.\footnote{To understand the trend with $\l$, Figures \ref{fig.EpolEtor} and  \ref{fig.fpol} show the distributions up to $\l=15$. We remind the reader that the other figures in this paper use $l \leq 5$.} Note that the toroidal energy is much lower than the poloidal energy during the whole cycle 24 (here, represented by the three plotted CRs). We do not show the distribution of total energies, as those are very similar in magnitude and in behaviour to the distributions of poloidal energies. The cumulative energies (left panels), as expected, increase with $\lmax$. The increase {in poloidal and toroidal energies} is steeper at maximum SSN (c.f., black versus blue/red curves), indicative of higher magnetic field strengths at this epoch of the solar cycle. {The magnetic energy in the axisymmetric component ($m=0$) is the largest at CR2202 (minimum). As we present in the next section, the solar magnetic field is more axisymmetric at epochs near minimum, and in particular when the sunspot number is in the decaying phase (i.e., at the end of the cycle). } Although the toroidal energies are at least one order of magnitude smaller than the poloidal energies over the sunspot cycle, the toroidal energy undergoes a sudden increase at $\lmax \geq 10$ at maximum activity. This is due to a high toroidal energy at $\l =10 $ (middle right panel). We believe this indicates that the toroidal field  comes from the bipoles -- the small-scale structures -- that become visible at higher $\ell$ degrees.

The panels on the right (Figure \ref{fig.EpolEtor}) also show that the poloidal (and total) magnetic energy of the dipole ($\l=1$) is smaller at maximum (black) than at minima (red, blue). The opposite holds for $\l \geq 2$ (quadrupole and above degrees), when the magnetic energy at a given $\l$ is larger during maximum than during minima. This is consistent with \citet{2012ApJ...757...96D} who showed that the dipole component of the radial magnetic field of the Sun is anti-correlated to sunspot number (smallest intensity at solar maximum). At minimum phases (red, blue), we see a zig-zag in the distribution of $\langle B^2_{{\rm pol},\ell}\rangle$, where even-$\l$ degrees have smallest energies in poloidal field,  when compared to adjacent odd-$\l$ degrees. In other words, at SSN minimum, the quadrupole ($\ell=2$) has the smallest poloidal energy compared to the dipole ($\ell=1$) or the octupole ($\ell=3$). The same is then true for most of the odd $\ell$ degrees. For the red curve, the `zig-zag' trend continues for higher $\l$, at least up to $\l=15$. \citet{2012ApJ...757...96D} already reported opposite behaviours in the energies between the $\ell=1$ and $\ell=2$ components of the line-of-sight magnetic field (at minimum, the dipolar energy is larger than the quadrupolar and vice-versa at maximum). Given that the line-of-sight component makes up for most of the poloidal field, what we seen in  Figure  \ref{fig.EpolEtor} for  $\ell=1$ and $\ell=2$ is the same as discussed in  \citet{2012ApJ...757...96D}.We note again that the energies per degree in the phases representative of SSN minima are considerably smaller than the energies in the phase representative of SSN maximum. 
 
The zig-zag trend described for the poloidal energy per degree is inverted for the toroidal energy (middle right panel). In the distribution of $\langle B^2_{{\rm tor},\ell}\rangle$ with $\l$, the even-$\l$ degrees have higher energies in toroidal field compared to adjacent odd-$\l$ degrees. For example, the toroidal energy of the quadrupole ($\l=2$) is the largest  compared to the toroidal energies in the dipole ($\l=1$) or in the octupole ($\l=3$). Likewise,  the `zig-zag' trend seems to continue for higher $\l$, at least up to $\l=15$ in the red curve.
Note also that the contribution of $\l=2$ to the toroidal energy is quite significant in the CR phases representative of either SSN minimum or maximum. At maximum, we see a significant increase in $\langle B^2_{{\rm tor},\ell}\rangle$ for $\l =10$, which, as mentioned before, we believe to be linked to the fact that the toroidal field is created by bipoles that  start to be visible at higher $\l$. The fact that we see opposite trends in the distribution of $\langle B^2_{{\rm pol},\ell}\rangle$ and $\langle B^2_{{\rm tor},\ell}\rangle$ is linked to the sun's magnetic field polarity, which is reversed across the equator. {Mathematically, this can be seen in Equations (\ref{eq.br}) and (\ref{eq.bphi}). For the solar magnetic field, most of its poloidal field is contained in the radial component, while the toroidal field is mainly stored in the azimuthal component. The radial poloidal field depends on $P\lm$ (Equation \ref{eq.br}) and the azimuthal toroidal field depends on ${\dif \plm}/{\dif \theta}$   (Equation \ref{eq.bphi}, term that contains $\gamma\lm$). The properties of symmetry about the equator for $P\lm$ and ${\dif \plm}/{\dif \theta}$ depend on the sum $(\ell+m)$. For odd $(\ell+m)$, $P\lm$ is antisymmetric and ${\dif \plm}/{\dif \theta}$ is symmetric about the equator. On the other hand, for even $(\ell+m)$, $P\lm$ is symmetric and ${\dif \plm}/{\dif \theta}$ is antisymmetric. Therefore, the equatorial polarity switches observed in the solar magnetic field (i.e., antisymmetry about the equator) require odd $(\ell+m)$ in the poloidal part (which depends on $P\lm$) and even $(\ell+m)$ in the toroidal part (which depends on ${\dif \plm}/{\dif \theta}$). 
 The inset in Figure 2 of \citet{2017MNRAS.466L..24L} provides a visualisation of this for the case of axisymmetric ($m=0$) dipolar and quadrupolar fields. These symmetry properties would lead to local maxima in poloidal fields with odd $(\ell+m)$ (and correspondingly local minima in toroidal fields) and to local minima in poloidal fields with even $(\ell+m)$ (and correspondingly local maxima in toroidal fields). The reason why the zig-zag is not very pronounced for larger $\ell > 4$ (especially in blue, red curves) could be because, to obtain $\langle B^2_{{\rm pol},\ell}\rangle$ and $\langle B^2_{{\rm tor},\ell}\rangle$, we already summed over $m \leq \ell$.} 

{A zig-zag pattern is also seen in the axisymmetric component of the solar magnetic field (bottom right panel). At minimum phases (red, blue), the trend resembles that of the poloidal energy (top right panel): in general, even-$\l$ degrees have smallest energies in axisymmetric field, when compared to adjacent odd-$\l$ degrees. In other words, at SSN minimum, the dipole ($\ell=1$) or the octupole ($\ell=3$) have larger axisymmetric energies compared to  the quadrupole ($\ell=2$). At maximum phase (black), on the other hand, the field is more axisymmetric at even $\ell$ degrees than adjacent odd $\ell$ degrees, but for $\ell \geq 7$, the zig-zag trend switches behaviour with more axisymmetric energies at odd $\ell$. }

{Figure \ref{fig.fpol} shows the ratio between toroidal and total energies of the large-scale field of the Sun (up to $l_{\rm max}=15$) for all the CRs available in this study. The top panel shows the ratio between the toroidal and total energy at a given $\ell$ degree ($f_{{\rm tor}, \ell}=  {\langle B_{{\rm tor},\ell}^2 \rangle}/{\langle B_{\ell}^2 \rangle}$), while the bottom panel represents the ratio between the toroidal and total cumulative energies ($f_{\rm tor}={\langle B_{{\rm tor},\Sigma}^2 \rangle}/{\langle B_{\Sigma}^2 \rangle}$)}. We see that the toroidal fraction at $\ell=2$ is high and usually higher than at $\l=1$, which indicates a more toroidal field in the quadrupole at most epochs during the solar activity cycle 24.  The exception is between CR2125 -- CR2150 (around maximum in SSN), where we also see a significant contribution of toroidal energy at $\ell =1$. The other interesting feature we observe in  Figure \ref{fig.fpol} is the vertical stripes in the toroidal fractions of the cumulative energies (bottom panel). {Similar vertical stripes also appear in the toroidal fraction of energies in individual $\ell$ degrees (top panel). Together, these features indicate} that changes in the toroidal energies occur simultaneously at all $\l$'s, i.e., there is no time delay in the increase of the toroidal energies at different components of the field. If the toroidal energy were first to appear in the small-scale structure and propagate with time towards large-scale structures, we would see a drift in toroidal fraction at each degree  from high-$\l$'s towards low-$\l$'s as a function of time, forming diagonal stripes in the diagram shown in the top panel.

\section{Evolution of the solar large-scale field through cycle 24}\label{sec.evolution}

\begin{figure*}
	\includegraphics[width=0.48\textwidth]{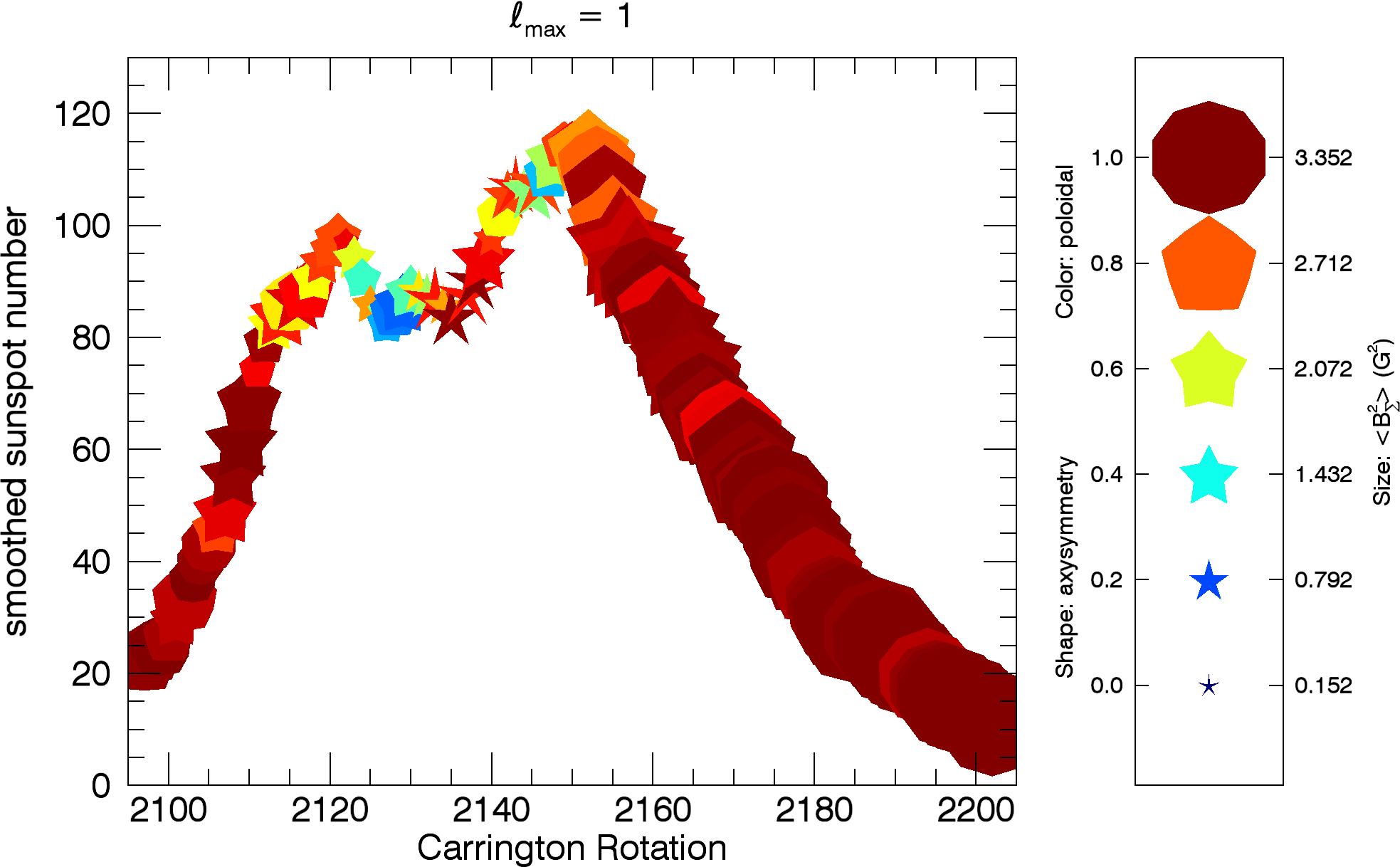}	\,\,\,\,\,\,
	\includegraphics[width=0.48\textwidth]{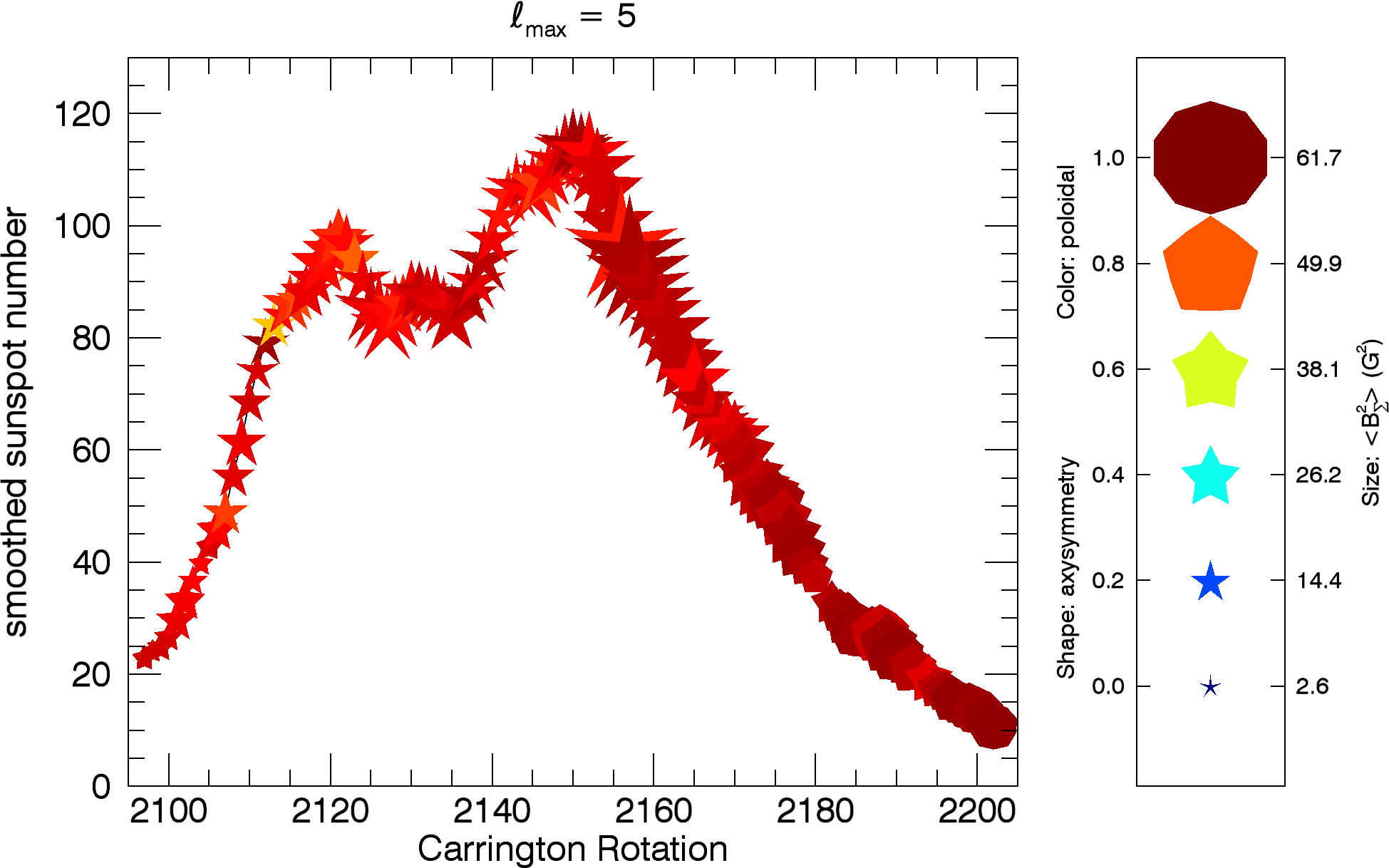}
\caption{Sunspot number (smoothed) as a function of CR for the solar cycle. The magnetic field reconstruction is done up to $\lmax=1$, i.e., dipole only (left) and $\lmax=5$ (right). Symbol sizes are proportional to magnetic energy $\langle B^2_\Sigma \rangle$, their colours indicate the fractional poloidal energy (ranging from deep red for purely poloidal field $f_{\rm pol}=1$ to blue for purely toroidal field $f_{\rm pol}=0$), and their shapes indicate the fraction of axisymmetric field (ranging from a decagon for purely axisymmetric field $f_{\rm axi}=1$ to a point-shaped star for $f_{\rm axi}=0$).}\label{fig.confusogram}
\end{figure*}

Figure \ref{fig.confusogram} shows the evolution of sunspot number over cycle 24. This diagram follows diagrams used in several ZDI studies \citep{2008MNRAS.390..567M,2009ARA&A..47..333D,2016MNRAS.455L..52V,2017MNRAS.472.4563M} -- they are useful to summarise magnetic field characteristics of a sample of magnetic maps. Each symbol carries information on the magnitude of the magnetic energy (symbol sizes are proportional to $\langle B^2_\Sigma \rangle$), on the fraction of poloidal energy (colours ranging from deep red for purely poloidal field $f_{\rm pol}=1$ to blue for purely toroidal field $f_{\rm pol}=0$), and the fraction of axisymmetric field (shapes ranging from a decagon for purely axisymmetric field $f_{\rm axi}=1$ to a point-shaped star for $f_{\rm axi}=0$).  The left panel of Figure \ref{fig.confusogram} shows the evolution of the dipolar field characteristics ($\lmax =1$) and the right panel shows the reconstruction done up to $\lmax=5$. 

The dipolar magnetic field, shown in the left panel, becomes less poloidal (blue/yellow/orange) near and around maximum SSN, between CR2120 and CR2150, when we also see frequent changes in symbol colours (variation in poloidal fractions). This is also when the dipole is less axisymmetric (more star shaped). Note the bigger symbols at minima, which are due to increase in dipolar field.

The `Sun-as-a-star' diagram is shown on the right panel ($\lmax=5$). There are several interesting characteristics that we can extract from this plot:
\begin{enumerate}
\item {\it Axisymmetry:} Firstly, the solar large-scale magnetic field is more axisymmetric  as it is near solar minima  (rounder symbols). This is more clearly seen around CR2200, but also visible at around CR2100. From CR2100 to CR2120, when the sunspot number is rising, the field axisymmetry decreases. After CR2160, when the sunspot number is decaying, axisymmetry increases. This implies that, for the same number of sunspots at different parts of cycle 24, $f_{\rm axi}$ will differ,  depending if the cycle is going towards maximum or towards minimum SSN. This cycle asymmetry is likely a result of the location of spots in the surface of the Sun as the cycle evolves. 
\item {\it Poloidal:} Secondly, the solar large-scale magnetic field is more poloidal as it approaches solar minima (darker symbols). Like in the axisymmetry discussion, the poloidal fraction seems to differ depending on whether the sunspot cycle is going towards maximum or towards minimum in SSN. For example,  in the decaying phase of the sunspot cycle, after CR2150,  $f_{\rm pol}$ increases, as the symbols become more dark red. In the rising phase of the sunspot cycle, between CR2100 and CR2120, $f_{\rm pol}$ seems to decrease overall, but this trend is not as clear as in the spot-decaying phase.
\item {\it Intensity/energy:} Thirdly, the overall magnetic energy is larger near sunspot maximum: bigger symbols around CR2150 (this is the opposite to what is seen in the left panel, where the dipole has larger intensity at minimum). The large-scale magnetic energy is higher in the decaying phase (bigger symbols). This is related to the formation of large-scale magnetic field. Prior to the  polar field reversal, there is significant cancellation between decaying flux of active regions and polar field. But after the polar field reversal, the following polarity flux contributes to strengthening up the polar field. Lehmann et al (in prep b) reported the same trend in the flux emergence simulations of solar cycle 23. 
\end{enumerate}

\begin{figure}
	\includegraphics[width=\columnwidth]{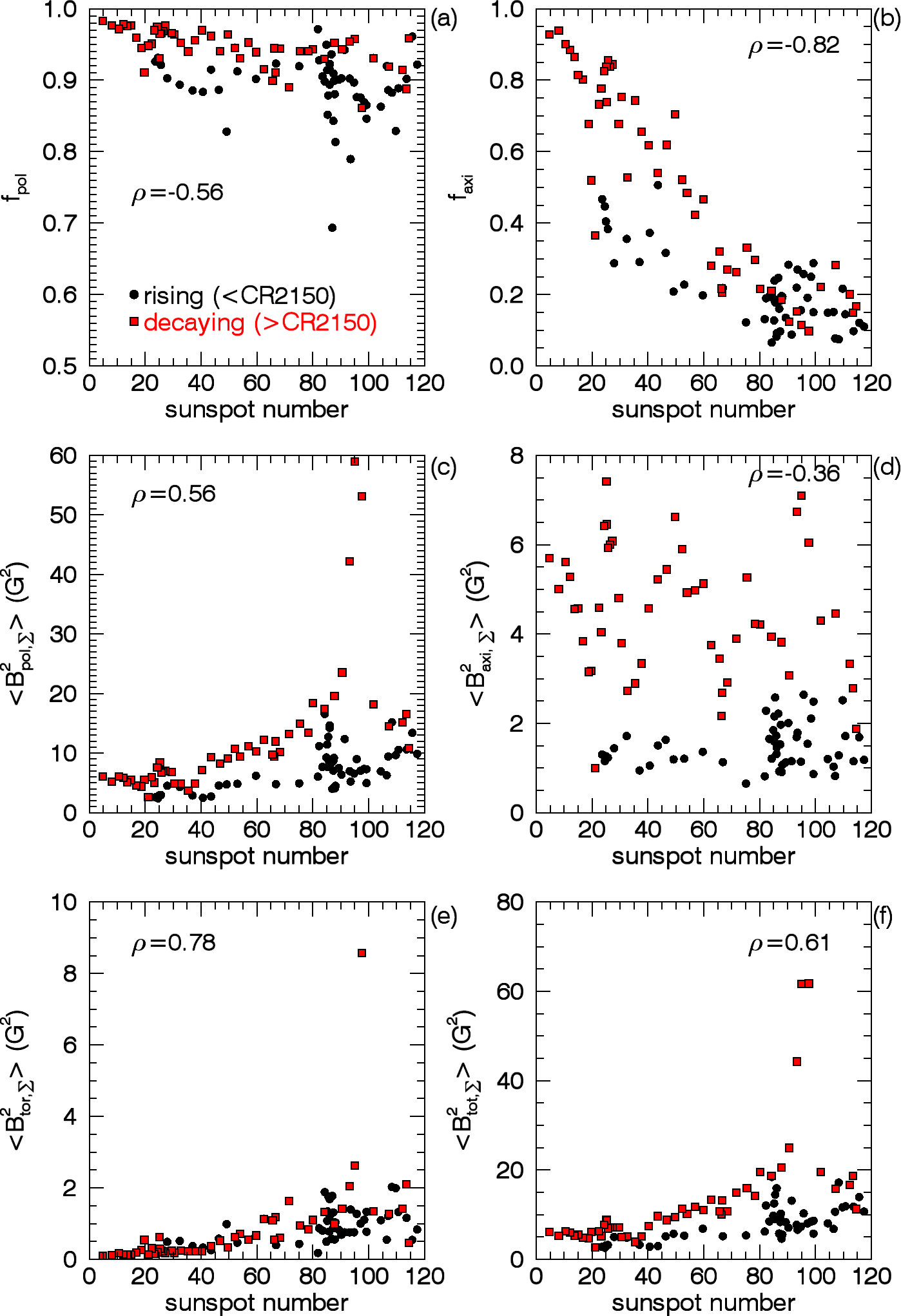}
	\caption{Properties of the large-scale ($\lmax=5$) magnetic field during cycle 24: (a) fraction of poloidal energy, (b) fraction of axisymmetric energy ($m=0$), (c) poloidal energy, (d) axisymmetric energy ($m=0$), (e) toroidal energy and (f) total magnetic energy (poloidal plus toroidal energies). Black circles (red squares) represent the phases of the cycle in which the number of sunspots is rising (decaying). We place this boundary at CR2150. Note that overall the red squares lie above black circles, i.e., overall quoted magnetic properties are larger in the decaying phase of the cycle than in the rising phase. {The  Spearman's rank correlation $\rho$ is shown in each panel (correlation considers circles and squares).}}\label{fig.magProperties}
\end{figure}

Because of some magnetic field properties seem to depend on whether the activity cycle is going towards maximum or towards minimum in SSN, we present in Figure \ref{fig.magProperties} several magnetic properties as a function of SSN. Red squares (black circles) represent phases of the solar cycle with decaying (rising) number of sunspots. We set the boundary between rising/decaying phases at CR2150.  Indeed, all the panels in Figure \ref{fig.magProperties} show red squares lying above  black circles, indicative of higher poloidal fraction, axisymmetric fraction and poloidal, toroidal and total energies that are overall larger in the decaying phase of the cycle than in the rising phase. The  asymmetry in the magnetic properties of the solar sunspot cycle is likely linked to the  short rise and a long decay phase of sunspot numbers \citep[e.g.,][]{2010LRSP....7....1H}. When we are in the rising phase of sunspot numbers, the magnetic field of decaying active regions is transported to the poles,  cancelling out the polar field. The polar field is close to zero around maximum of sunspot. About one or two years after the maximum of the  solar cycle, the polar field reverses its sign and is rebuilt (getting stronger and stronger) during the declining phase of cycle. I.e., we start to see the build up of polar fields, while bipoles are being removed. 

There are two tight correlations worth mentioning. The first one is between the toroidal energy and sunspot number (Figure \ref{fig.magProperties}e, Spearman's rank correlation is $\rho=0.78$, with a non-null probability $\ll10^{-10}$). This indicates that sunspots are major contributors to the toroidal energy. This supports the discussion in Section \ref{sec.reversals}, in which the $B_\varphi$ component does not cancel out with spatial averaging (unlike $B_r$). The second one is the anti-correlation between $f_{\rm axi}$ and sunspot number (Figure \ref{fig.magProperties}b), which shows that the larger the number of sunspots, the less axisymmetric  the field is ($\rho=-0.82$ and non-null probability $\ll10^{-10}$, see  \citealt{2013ApJ...768..162P} in the context of line-of-sight fields). In spite of this tight relation, a  large spread is seen in the axisymmetric energy, which ranges from 1 to 8 G$^2$ (panel d). However, the total magnetic energy increases with sunspot number (panel f), explaining why the axisymmetric fraction (ratio between panels d and f) decreases with SSN.

\section{Placing the solar large-scale field in the context of stars}\label{sec.zdi}
\begin{figure}
	\includegraphics[width=0.47\textwidth]{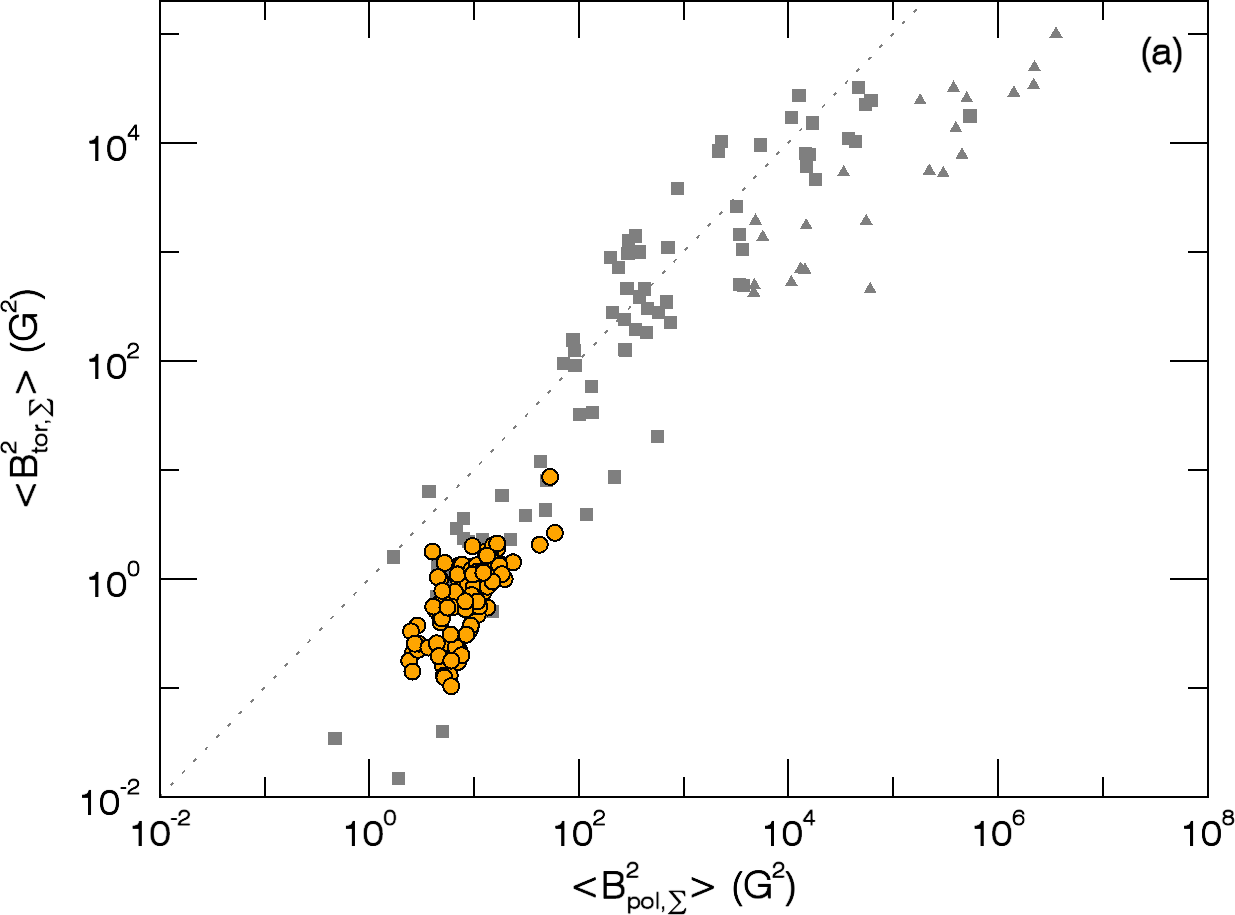}\\	
	\includegraphics[width=0.47\textwidth]{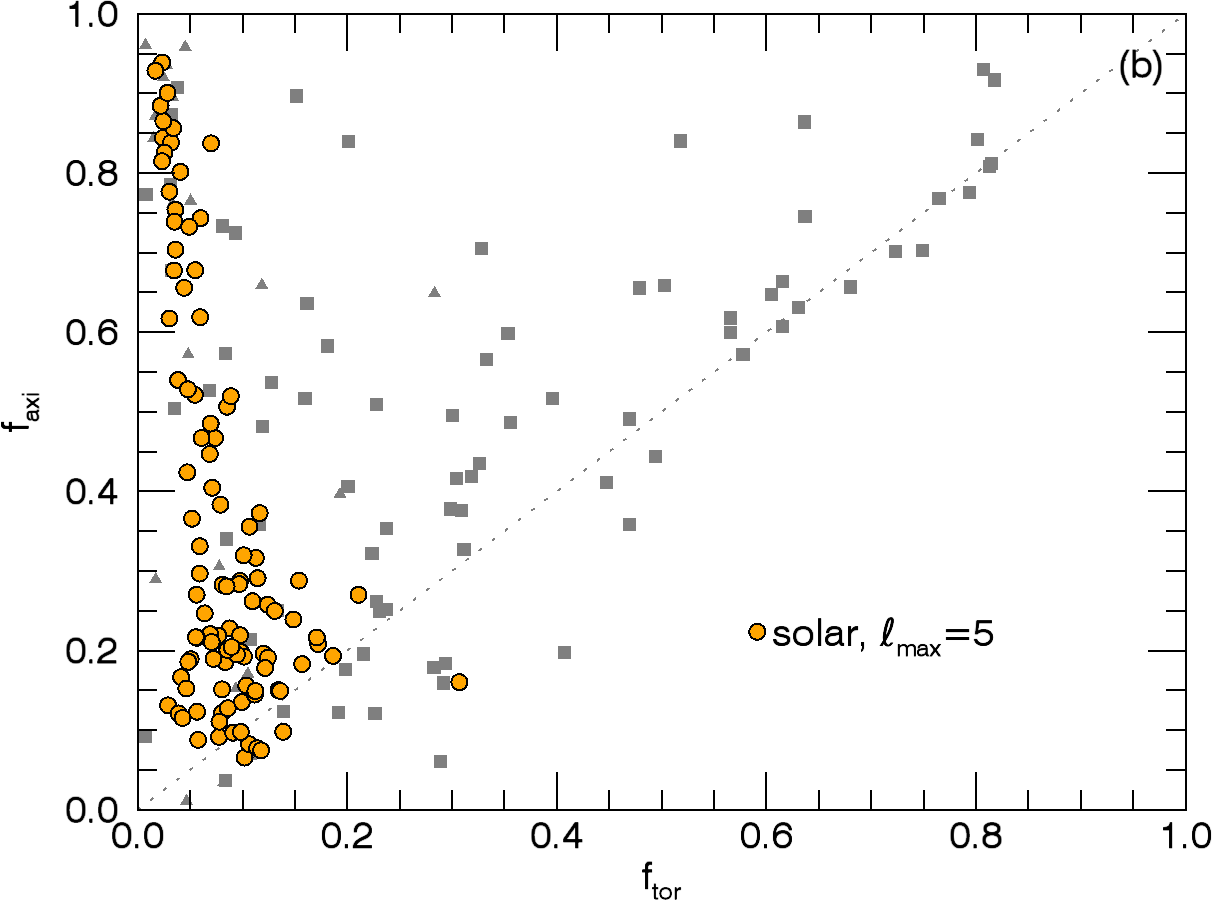}
	\caption{(a) The toroidal magnetic energies are correlated to the poloidal energies. The solar points are shown in colour orange for $\lmax = 5$. Stars with reconstructed magnetic fields are shown as grey symbols: triangles for masses $<0.5M_\odot$ and squares for $\geq 0.5M_\odot$. (b) The fraction of magnetic energy contained in the axisymmetric component of the field ($m=0$) against the fraction of the magnetic energy contained in the toroidal field. Symbols are as in panel (a).}\label{fig.zdicomparison}
\end{figure}

By reconstructing the solar magnetic vector field up to $\lmax = 5$, we can more directly compare the solar magnetic properties with the magnetic properties of low-mass stars. From a sample of low-mass stars with reconstructed magnetic fields using the ZDI technique, \citet{2015MNRAS.453.4301S} showed that the toroidal magnetic energies are correlated to the poloidal energies. For stars with masses $<0.5~M_\odot$, $\etor \propto \epol ^{0.72}$, and for masses $>0.5~M_\odot$, $\etor \propto \epol ^{1.25}$. These points are shown as grey symbols in Figure \ref{fig.zdicomparison}a: squares (triangles) denote  stars with masses above (below) $0.5~M_\odot$. Figure \ref{fig.zdicomparison}a also shows the solar magnetic field energies through cycle 24, reconstructed up to degree $\lmax = 5$ (orange). The solar points merge very smoothly with the other stars and fall on the bottom part of the figure, towards low energies. The axes span nearly 8 orders of magnitude in toroidal and poloidal energies. If we focus on the orange solar points, we can infer the expected variability of magnetic energies of solar-like stars through activity cycles. This cycle variability increases the spread in magnetism relations \citep[e.g.,][]{2014MNRAS.441.2361V, 2016MNRAS.462.4442S}. Fitting a power-law through the solar ($\lmax=5$) and ZDI points together, we get a slightly steeper correlation $\etor \propto \epol ^{1.38\pm 0.04}$ than from the ZDI stars only (power law index of $1.26 \pm 0.05$), with $\rho=0.85$. The correlation with only the solar points at $\lmax=5$ provides a power-law index of $1.38 \pm 0.09$.

Following on another result from \citet{2015MNRAS.453.4301S}, Figure \ref{fig.zdicomparison}b shows the fraction of magnetic energy contained in the axisymmetric component of the field ($m=0$) against the fraction of the magnetic energy contained in the toroidal field. Symbols are as in panel (a).  \citet{2015MNRAS.453.4301S} noted a  dearth of toroidal non-axisymmetric stars, in the lower-right corner of the plot. By adding the solar points to Figure \ref{fig.zdicomparison}b, the trend  noticed by \citet{2015MNRAS.453.4301S} remains. Again, focusing only on the solar points (orange), we note the variation during the solar cycle in the $f_{\rm axi}$ versus $f_{\rm tor}$ plane . As we move towards the end of cycle 24, the orange points start to populate  the top left corner of the diagram (increasing $f_{\rm axi}$ and decreasing $f_{\rm tor}$). The 3D non-potential flux transport simulations of the solar cycle 23 populate the same parameter range as the observations of cycle 24 (Lehmann et al., in prep b).

Another technique used to measure the magnetic fields of stars is the Zeeman broadening (ZB) technique \citep[e.g.][]{2001ASPC..223..292S,1999ApJ...516..900J,2009ApJ...692..538R}. This is a complementary technique to ZDI: while ZDI is not sensitive to magnetic fluxes in small scales, ZB is able to measure the total (small and large scales) unsigned magnetic flux of a star. However, ZB cannot recover field geometry as ZDI does. For a sample of M dwarfs, \citet{2009A&A...496..787R} and \citet{2010MNRAS.407.2269M} showed that the ratio between the average magnetic energy recovered in ZDI ($\bv$) and the average magnetic energy recovered in ZB  ($\bi$)  ranges from 6 to 14\%. Here, we use the high-resolution solar magnetic maps to calculate the unsigned magnetic field strength in the total field and we adopt this value as the equivalent of $\bi$. We use the total magnetic field calculated at $\lmax=5$ as an equivalent to $\bv$. With these derived values, we compute the ratio $\bv/\bi$ over  solar cycle 24.  We found that this ratio is below 27\% during the whole cycle 24 and most often between 10 and 20\%, indicating that only 10 to 20\% of the total magnetic field of the Sun would be recovered in ZDI observations. This is consistent with the results found by Lehmann et al. (in prep. a) when applying ZDI to solar-like star simulations.
 The number of objects with magnetic field measurements using both the Zeeman broadening technique and and ZDI technique is going to increase manyfold with SPIRou, a near-infrared spectropolarimeter that was recently installed in CFHT.   

\begin{figure}
	\includegraphics[width=0.47\textwidth]{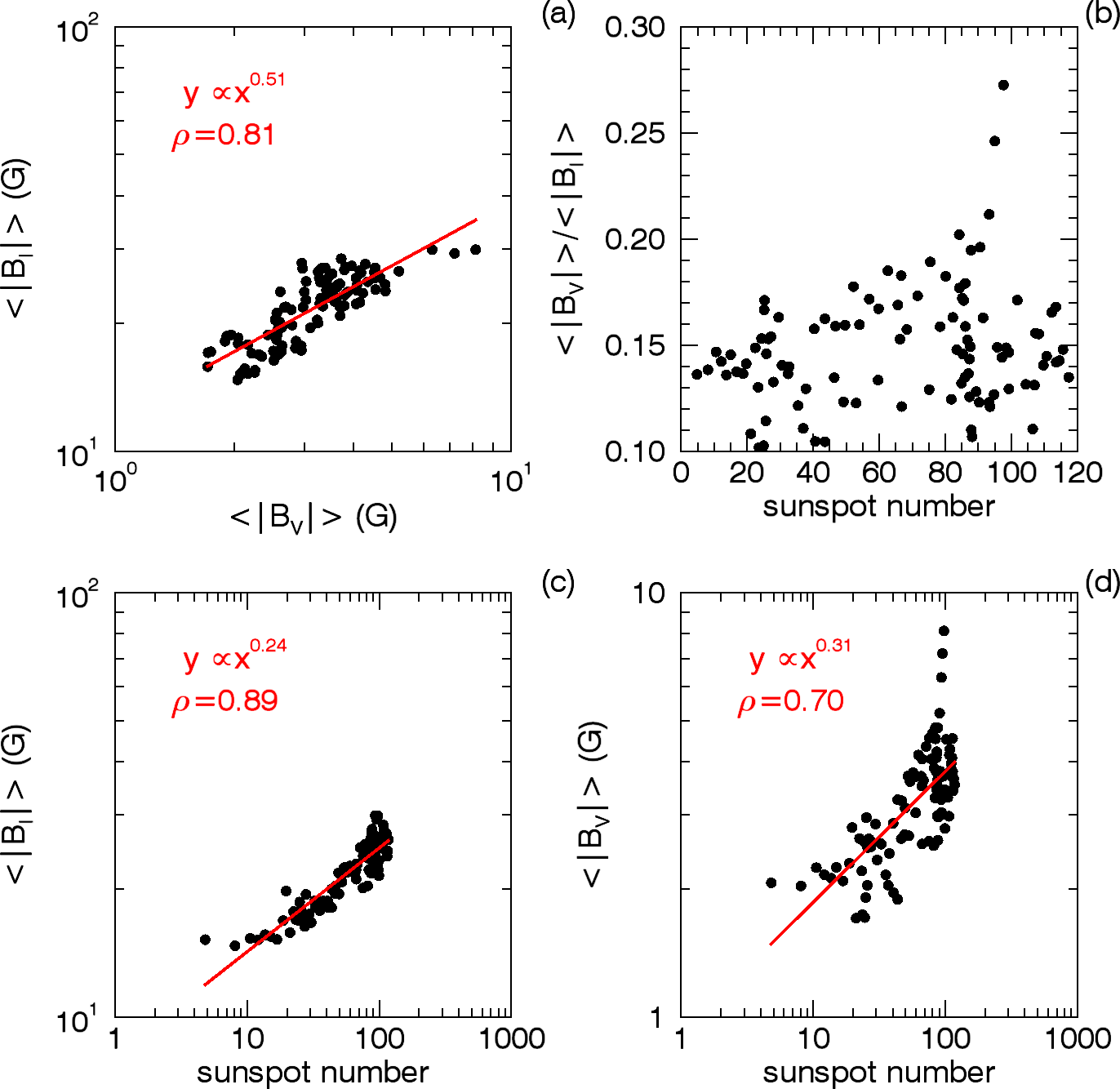}
\caption{Correlations between the average magnetic energy recovered by ZDI $\bv$ and the average magnetic energy recovered in Zeeman broadening $\bi$. We use the total magnetic field calculated at $\lmax=5$ as an equivalent to $\bv$ and the high-resolution (observed) solar magnetic maps as the equivalent to $\bi$. Power-law fits are shown for all panels (red), except b, along with the Pearson correlation coefficient $\rho$.}\label{fig.BiBv}
\end{figure}

We also investigate the presence of correlations between  $\bv$ and  $\bi$ (Figure \ref{fig.BiBv}a),   $\bv/\bi$ and SSN (\ref{fig.BiBv}b)  $\bi$ and SSN (\ref{fig.BiBv}c) and $\bv$ and SSN (\ref{fig.BiBv}d).  Among all these, we found a tight correlation between $\bi$ and SSN (Spearman's rank coefficient $\rho=0.89$, null-probability $\ll 10^{-10}$). We also found that $\bi \propto \bv^{1/2}$ ($\rho=0.81$, null-probability $\ll 10^{-10}$), which implies that an increase of a factor of $~10$ in $\bi$, increase about a factor of $3$ in $\bv$.  

\section{Summary and Conclusions}\label{sec.conclusions}
In this study, we decomposed the synoptic maps of the solar vector magnetic field in terms of spherical harmonics. We used the vector synoptic maps from HMI \citep{liu}, which are available since CR2097 and currently extend up to CR2202. During this period, the Sun evolved from nearly a minimum in sunspot number (SSN), passing through a maximum at around CR2150 and then nearly reaching the consecutive minimum at the end of the observing period (Figure \ref{fig.SN}). Although this does not span a full magnetic cycle (22 yr), it nearly spans a full activity cycle (11yr).

Once the vector synoptic maps were decomposed using spherical harmonics, we used the derived coefficients to reconstruct the solar vector field. This method was described in \citet{2016MNRAS.459.1533V}. Because we are interested in studying the Sun as a star, we limit the reconstruction to a maximum spherical harmonics degree $\lmax=5$. Our description is entirely consistent with the description adopted in several ZDI studies \citep[e.g.][]{2006MNRAS.370..629D}. By filtering out the spherical harmonics with high degrees, our method allows one to transform high-resolution vector synoptic maps of the Sun into maps with similar resolution to those derived in stellar studies, allowing, therefore, their direct comparison. It has been shown that the spherical harmonics decomposition method provides a good proxy for what ZDI  recovers for solar-like stars (Lehmann et al., in prep. a). 

We showed in Figure \ref{fig.sunmap}  the large-scale vector field of the Sun reconstruct up to $\l_{\rm max}=5$, which is a typical maximum degree achieved in ZDI studies. Although we have analysed 104 CRs, we only presented the field reconstruction for three of them: one representative of a minimum of SSN (CR2097), one representative of a maximum (CR2150) and then another one representative of a following minimum (CR2202). A polarity reversal is immediately seen in the radial and meridional components of the large-scale solar field, but not in the azimuthal component, which remains negative (positive) in the northern (southern) hemisphere during the entire timespan of our maps (Figure \ref{fig.butterfly}). The lack of reversal in the $\varphi$ component is due to our reduced timespan of maps: new bipoles of different polarities will start to emerge in the next sunspot cycle and consequently we expect $B_\varphi$ to reverse its polarity.

The radial or line-of-sight large-scale field of the Sun has been investigated in several works \citep[e.g.][]{2012ApJ...757...96D, 2013ApJ...768..162P}. Our work focused on the decomposition of the solar field into poloidal and toroidal components, which was not previously studied. We showed that the solar field has been mainly poloidal during cycle 24, with $\gtrsim 70\%$ poloidal energy for $\lmax =5$ (the average through the analysed timespan is $90\%$, see Figure \ref{fig.magProperties}a). When analysing poloidal and toroidal energies at each spherical harmonic degree, we found, in general, that the toroidal energies have local maxima in even $\l$ degrees, while the poloidal energies have local maxima in odd $\l$ degrees  (see Figures \ref{fig.EpolEtor} and \ref{fig.fpol}). For example, while the toroidal energy of the quadrupole ($\l=2$) is larger than that of the dipole ($\l=1$) or octupole ($\l=3$), for the poloidal energy, the inverse is true   (see Figure \ref{fig.EpolEtor}, right panels). This happens because the sun's magnetic polarity switches across the equator, which enhances even toroidal degrees and odd poloidal degrees \citep{2017MNRAS.466L..24L, 2018MNRAS.tmp.1193L}. This creates a `zig-zag' distribution of toroidal or poloidal energies per degree as a function of $\l$. In general the toroidal energy of the quadrupole is larger than that of the dipole or octupole throughout cycle 24, being the largest (or one of the largest) contributors to the reduction in poloidal field fraction through the cycle (Figure \ref{fig.fpol}). 

We investigated the magnetic properties of the `Sun-as-a-star'  ($\lmax=5$) as a function of time (Figure \ref{fig.confusogram}). We found that the solar large-scale magnetic field is more axisymmetric and more poloidal as it is near solar minima. (In our study, an axisymmetric field is defined as having spherical harmonics mode $m=0$.) The overall magnetic energy is larger near sunspot maximum. We also noticed that some magnetic field properties depend on whether the cycle is going towards maximum or towards minimum in SSN. All the panels in Figure \ref{fig.magProperties} show that  poloidal fraction, axisymmetric fraction and poloidal, toroidal and total energies are overall larger in the decaying phase of  cycle 24 than in the rising phase.
We found a tight anti-correlation (Spearman's rank correlation is $\rho=-0.82$) between axisymmetric energy fraction and sunspot number (Figure \ref{fig.magProperties}b), which means that the larger the number of spots (near maximum), the smaller the fraction of axisymmetric fields. We also found that toroidal energy is correlated to sunspot number ($\rho=0.78$), which might indicate that sunspots are major contributors to the toroidal large-scale energy of the Sun. 

The ZDI technique (Stokes V) is only able to capture the large-scale surface magnetic fields of stars,  while the small-scale field is better captured by Zeeman broadening measurements of unpolarised (Stokes I) lines. As an analogy to the latter method, we used the high-resolution observed magnetic field of the Sun to compute the unsigned total magnetic field of the Sun ($\bi$) and we used the $\lmax=5$ reconstructed magnetic field to compute the unsigned large-scale magnetic field of the Sun ($\bv$). We found that $\bv/\bi$ is often between 10 and 20\% during cycle 24, indicating that only 10 to 20\% of the total magnetic field of the Sun would be recovered in ZDI observations of the Sun-as-a-star (Figure \ref{fig.BiBv}).

We finally investigated how the properties of the large-scale vector field of the Sun compare to magnetic field measurements of a ZDI stellar sample. The sample used in this comparison is the same as that presented in \citet{2014MNRAS.441.2361V} and \citet{2015MNRAS.453.4301S}. It is striking to see how smoothly the solar points fit in the stellar sample (Figure \ref{fig.zdicomparison}). Together, the solar and stellar points span nearly 8 orders of magnitude in toroidal and poloidal energies, with a tight  ($\rho=0.85$) correlation $\etor \propto \epol ^{1.38 \pm 0.04}$ (Figure \ref{fig.zdicomparison}a). This is consistent with the correlation found for the stellar sample: $\etor \propto \epol ^{1.26 \pm 0.05}$. In Figure \ref{fig.zdicomparison}b, we  investigated how the solar points are distributed in the $f_{\rm axi}$ vs $f_{\rm tor}$ plane and found that they follow very closely the trend found by \citet{2015MNRAS.453.4301S}, who reported a dearth of toroidal non-axisymmetric stars. It is again striking how well the Sun-as-a-star points follow the trends of ZDI-derived magnetism in stars.

\section*{Acknowledgements}

The sunspot number used in this work are from WDC-SILSO, Royal Observatory of Belgium, Brussels. JSOC/HMI maps: Courtesy of NASA/SDO and the AIA, EVE, and HMI science teams. LTL acknowledges support from the Scottish Universities Physics Alliance (SUPA) prize studentship and the University of St Andrews Higgs studentship. AAP acknowledges partial support by NASA grant NNX15AN43G. National Solar Observatory is operated by the Association of Universities for Research in Astronomy (AURA)  Inc., under a Cooperative agreement with the National Science Foundation. 

\appendix
\section{Spherical harmonic coefficients of maps in Figure 2} \label{sec.coeff}
Tables \ref{table.coeffs-2097}, \ref{table.coeffs-2150}, \ref{table.coeffs-2202}  show the coefficients for the first five  degrees derived from the spherical harmonics decomposition of the solar vector field at CR2097 (near minimum of SSN) and CR2150 (at maximum of SSN), and CR2202 (near minimum of SSN), respectively. To reconstruct the large-scale magnetic field shown in Figure \ref{fig.sunmap}, Equations (\ref{eq.br}) to (\ref{eq.bphi})  should be used.

\begin{table} \begin{center}
\caption{Spherical harmonic coefficients derived in the decomposition of the magnetic field distribution of the solar synoptic map of the vector field up to degree $\lmax=5$. Coefficients are for CR2097, near minimum of sunspot number (left column of Figure \ref{fig.sunmap}). Each set of coefficients is split between its real and imaginary parts. \label{table.coeffs-2097}}
\begin{tabular}{rrrrrrrr}
\hline
$\l$ & $m$ & \multicolumn{2}{c}{${\alpha\lm} $ (G)}   & \multicolumn{2}{c}{${\beta\lm} $ (G)}   & \multicolumn{2}{c}{${\gamma\lm} $ (G)}   \\
& & \multicolumn{1}{c}{Re}  & \multicolumn{1}{c}{Im}  & \multicolumn{1}{c}{Re}  & \multicolumn{1}{c}{Im}  & \multicolumn{1}{c}{Re}  & \multicolumn{1}{c}{Im}  \\
 \hline \hline
 $    1$ & $    0$ & $    -2.710$ & $    0$ & $    -0.044$ & $    0$ & $     0.396$ & $    0 $ \\
&$    1$ & $    -0.711$ & $     1.564$ & $     0.363$ & $    -0.549$ & $    -0.003$ & $    -0.590 $ \\
 $    2$ & $    0$ & $     0.422$ & $    0$ & $    -0.028$ & $    0$ & $     1.311$ & $    0 $ \\
&$    1$ & $    -1.714$ & $     0.436$ & $    -0.331$ & $    -0.193$ & $    -0.303$ & $    -0.406 $ \\
&$    2$ & $    -0.211$ & $     1.129$ & $    -0.140$ & $     0.091$ & $    -0.500$ & $    -0.175 $ \\
 $    3$ & $    0$ & $    -2.252$ & $    0$ & $    -0.483$ & $    0$ & $     0.220$ & $    0 $ \\
&$    1$ & $    -1.648$ & $     1.663$ & $     0.240$ & $     0.414$ & $    -0.084$ & $    -0.557 $ \\
&$    2$ & $    -0.689$ & $     0.976$ & $     0.080$ & $     0.133$ & $    -0.143$ & $    -0.475 $ \\
&$    3$ & $     0.711$ & $     2.266$ & $     0.120$ & $     0.075$ & $    -0.057$ & $     0.555 $ \\
 $    4$ & $    0$ & $     0.148$ & $    0$ & $     0.014$ & $    0$ & $    -0.262$ & $    0 $ \\
&$    1$ & $    -0.959$ & $    -0.990$ & $    -0.386$ & $     0.297$ & $    -0.015$ & $    -0.085 $ \\
&$    2$ & $    -0.307$ & $     0.448$ & $     0.066$ & $     0.015$ & $     0.173$ & $     0.035 $ \\
&$    3$ & $     1.145$ & $     0.885$ & $    -0.267$ & $    -0.060$ & $    -0.877$ & $     0.050 $ \\
&$    4$ & $    -1.310$ & $    -0.828$ & $     0.262$ & $     0.306$ & $     0.110$ & $    -0.440 $ \\
 $    5$ & $    0$ & $    -1.488$ & $    0$ & $    -0.493$ & $    0$ & $     0.081$ & $    0 $ \\
&$    1$ & $    -0.901$ & $    -0.703$ & $    -0.559$ & $    -0.019$ & $     0.340$ & $     0.325 $ \\
&$    2$ & $     0.079$ & $     0.065$ & $     0.010$ & $     0.019$ & $     0.104$ & $    -0.027 $ \\
&$    3$ & $    -0.071$ & $     1.973$ & $    -0.143$ & $     0.431$ & $    -0.283$ & $    -0.263 $ \\
&$    4$ & $    -0.915$ & $    -0.074$ & $     0.083$ & $     0.253$ & $     0.327$ & $     0.332 $ \\
&$    5$ & $    -0.419$ & $     0.605$ & $     0.171$ & $    -0.130$ & $    -0.172$ & $     0.072 $ \\
\hline
\end{tabular}
\end{center}
\end{table}

\begin{table}
\begin{center}
\caption{Same as Table \ref{table.coeffs-2097}, but for CR2150, at maximum of SSN (middle column of Figure \ref{fig.sunmap}). \label{table.coeffs-2150}}
\begin{tabular}{rrrrrrrrrr}
\hline $\l$ & $m$ & \multicolumn{2}{c}{${\alpha\lm} $ (G)}   & \multicolumn{2}{c}{${\beta\lm} $ (G)}   & \multicolumn{2}{c}{${\gamma\lm} $ (G)}   \\
& & \multicolumn{1}{c}{Re}  & \multicolumn{1}{c}{Im}  & \multicolumn{1}{c}{Re}  & \multicolumn{1}{c}{Im}  & \multicolumn{1}{c}{Re}  & \multicolumn{1}{c}{Im}  \\ \hline \hline
 $    1$ & $    0$ & $     1.537$ & $    0$ & $     0.431$ & $    0$ & $    -0.914$ & $    0 $ \\
&$    1$ & $     1.169$ & $     0.619$ & $     0.282$ & $    -0.445$ & $     0.492$ & $    -0.096 $ \\
 $    2$ & $    0$ & $    -1.644$ & $    0$ & $    -0.724$ & $    0$ & $     1.494$ & $    0 $ \\
&$    1$ & $     0.048$ & $    -0.626$ & $     0.334$ & $    -0.236$ & $     0.074$ & $    -0.160 $ \\
&$    2$ & $    -2.260$ & $    -3.376$ & $    -0.606$ & $     0.169$ & $    -1.170$ & $     0.892 $ \\
 $    3$ & $    0$ & $     0.206$ & $    0$ & $     0.811$ & $    0$ & $    -0.040$ & $    0 $ \\
&$    1$ & $    -0.621$ & $     0.313$ & $    -0.219$ & $    -0.135$ & $    -0.178$ & $     0.619 $ \\
&$    2$ & $     2.780$ & $     1.531$ & $     0.995$ & $     0.328$ & $     0.788$ & $    -1.134 $ \\
&$    3$ & $    -7.219$ & $     1.005$ & $    -0.400$ & $     1.050$ & $    -0.274$ & $     0.663 $ \\
 $    4$ & $    0$ & $     2.410$ & $    0$ & $    -0.450$ & $    0$ & $    -0.220$ & $    0 $ \\
&$    1$ & $    -0.627$ & $    -0.180$ & $    -0.134$ & $    -0.149$ & $    -0.146$ & $    -0.021 $ \\
&$    2$ & $    -1.043$ & $     1.312$ & $    -0.705$ & $    -0.775$ & $     0.488$ & $     0.426 $ \\
&$    3$ & $     0.768$ & $    -1.610$ & $    -0.088$ & $    -0.196$ & $    -0.022$ & $    -1.605 $ \\
&$    4$ & $    -5.082$ & $     6.991$ & $    -0.583$ & $     0.758$ & $     1.126$ & $    -0.162 $ \\
 $    5$ & $    0$ & $    -2.656$ & $    0$ & $    -0.685$ & $    0$ & $     0.239$ & $    0 $ \\
&$    1$ & $     0.365$ & $     0.173$ & $     0.644$ & $     0.278$ & $     0.239$ & $    -0.566 $ \\
&$    2$ & $    -3.359$ & $    -1.229$ & $    -0.458$ & $    -0.125$ & $    -0.463$ & $     0.394 $ \\
&$    3$ & $     0.886$ & $    -0.765$ & $    -0.932$ & $    -0.641$ & $     0.170$ & $    -0.045 $ \\
&$    4$ & $    -1.048$ & $    -3.592$ & $    -0.322$ & $    -0.414$ & $    -1.493$ & $    -0.255 $ \\
&$    5$ & $     8.493$ & $     5.290$ & $     0.650$ & $     1.038$ & $    -0.050$ & $    -0.060 $ \\
\hline
\end{tabular}
\end{center}
\end{table}

\begin{table}
\begin{center}
\caption{Same as Table \ref{table.coeffs-2097}, but for CR2202, near minimum of SSN (right column of Figure \ref{fig.sunmap}). \label{table.coeffs-2202}}
\begin{tabular}{rrrrrrrrrr}
\hline $\l$ & $m$ & \multicolumn{2}{c}{${\alpha\lm} $ (G)}   & \multicolumn{2}{c}{${\beta\lm} $ (G)}   & \multicolumn{2}{c}{${\gamma\lm} $ (G)}   \\
& & \multicolumn{1}{c}{Re}  & \multicolumn{1}{c}{Im}  & \multicolumn{1}{c}{Re}  & \multicolumn{1}{c}{Im}  & \multicolumn{1}{c}{Re}  & \multicolumn{1}{c}{Im}  \\ \hline \hline
 $    1$ & $    0$ & $     5.406$ & $    0$ & $     0.526$ & $    0$ & $    -0.120$ & $    0 $ \\
&$    1$ & $     0.181$ & $    -0.461$ & $     0.042$ & $    -0.139$ & $     0.304$ & $     0.190 $ \\
 $    2$ & $    0$ & $    -1.619$ & $    0$ & $    -0.260$ & $    0$ & $     0.793$ & $    0 $ \\
&$    1$ & $     1.152$ & $    -1.286$ & $    -0.044$ & $    -0.067$ & $    -0.160$ & $     0.214 $ \\
&$    2$ & $    -0.329$ & $     0.519$ & $     0.030$ & $     0.005$ & $    -0.255$ & $     0.212 $ \\
 $    3$ & $    0$ & $     5.224$ & $    0$ & $     1.358$ & $    0$ & $    -0.496$ & $    0 $ \\
&$    1$ & $     1.007$ & $     0.951$ & $     0.306$ & $    -0.389$ & $     0.088$ & $     0.394 $ \\
&$    2$ & $     0.152$ & $     0.197$ & $    -0.024$ & $    -0.133$ & $    -0.145$ & $    -0.079 $ \\
&$    3$ & $    -0.111$ & $    -0.594$ & $    -0.136$ & $     0.139$ & $     0.066$ & $    -0.027 $ \\
 $    4$ & $    0$ & $    -3.098$ & $    0$ & $    -0.651$ & $    0$ & $     0.496$ & $    0 $ \\
&$    1$ & $     0.839$ & $     0.431$ & $     0.077$ & $    -0.104$ & $    -0.356$ & $     0.208 $ \\
&$    2$ & $     0.233$ & $     0.248$ & $    -0.116$ & $     0.173$ & $    -0.087$ & $     0.248 $ \\
&$    3$ & $     0.173$ & $     0.015$ & $     0.040$ & $     0.182$ & $    -0.162$ & $     0.217 $ \\
&$    4$ & $    -0.208$ & $    -0.197$ & $    -0.074$ & $     0.023$ & $     0.143$ & $    -0.024 $ \\
 $    5$ & $    0$ & $     0.932$ & $    0$ & $     1.183$ & $    0$ & $    -0.413$ & $    0 $ \\
&$    1$ & $    -0.259$ & $     1.014$ & $     0.141$ & $     0.037$ & $    -0.093$ & $     0.026 $ \\
&$    2$ & $     0.965$ & $     0.258$ & $     0.152$ & $     0.157$ & $     0.143$ & $     0.075 $ \\
&$    3$ & $     0.493$ & $     0.387$ & $     0.084$ & $     0.147$ & $     0.225$ & $    -0.188 $ \\
&$    4$ & $     0.403$ & $     0.195$ & $     0.050$ & $    -0.055$ & $     0.054$ & $    -0.035 $ \\
&$    5$ & $     0.156$ & $    -0.265$ & $    -0.004$ & $    -0.044$ & $     0.082$ & $    -0.161 $ \\
\hline
\end{tabular}
\end{center}
\end{table}

\bsp
\label{lastpage}
\end{document}